\documentclass[11pt,a4paper,english,twoside]{article}

\usepackage{a4wide}
\usepackage{amssymb, amsmath}
\usepackage{graphicx}
\usepackage[all]{xy}
\usepackage{enumerate}
\usepackage{dsfont}
\usepackage{empheq}
\usepackage{cite}
\usepackage{float}
\usepackage[latin1]{inputenc}
\usepackage[T1]{fontenc}
\usepackage{hhline}
\usepackage{soul}
\usepackage[pdftex,hyperref,x11names]{xcolor}
\usepackage[pdftex,colorlinks=true,
pdfstartview=FitV,
pdfnewwindow=true,
linktoc = page,
linkcolor= RoyalBlue3,
citecolor= Red3,
urlcolor= RoyalBlue3,
hyperindex=true,
hyperfigures=false]{hyperref}
\usepackage{subcaption}

\newcommand{\beq}{\begin{equation}}
\newcommand{\eeq}{\end{equation}}
\def\bea#1\eea{\begin{align}#1\end{align}}
\def\beal#1\eeal{\begin{subequations}\begin{align}#1\end{align}\end{subequations}}
\newcommand{\nn}{\nonumber}
\newcommand{\w}{\wedge}

\newcommand{\eq}[1]{\begin{equation}#1\end{equation}}
\newcommand{\spl}[1]{\begin{split}#1\end{split}}

\def\del {\partial}
\def\d {{\rm d}}

\begin{document}
\numberwithin{equation}{section}

\begin{titlepage}
\begin{center}

\phantom{DRAFT}

{\LARGE \bf{Accelerated expansion of an open universe,}\\ \vspace{0.1in}\bf{ and string theory realizations}}\\

\vspace{2.2 cm} {\Large David Andriot$^{1}$, Dimitrios Tsimpis$^{2}$, Timm Wrase$^{3,1}$}\\
\vspace{0.9 cm} {\small\slshape $^1$ Laboratoire d'Annecy-le-Vieux de Physique Th\'eorique (LAPTh),\\
CNRS, Universit\'e Savoie Mont Blanc (USMB), UMR 5108,\\
9 Chemin de Bellevue, 74940 Annecy, France}\\
\vspace{0.2 cm} {\small\slshape $^2$ Institut de Physique des Deux Infinis de Lyon (iP2i),\\
Universit\'{e} de Lyon, UCBL, CNRS/IN2P3, UMR 5822,\\
4 rue Enrico Fermi, 69622 Villeurbanne Cedex, France}\\
\vspace{0.2 cm} {\small\slshape $^3$ Department of Physics, Lehigh University,\\
16 Memorial Drive East, Bethlehem, PA 18018, USA}\\
\vspace{0.5cm} {\upshape\ttfamily andriot@lapth.cnrs.fr; tsimpis@ipnl.in2p3.fr; timm.wrase@lehigh.edu}\\

\vspace{2.8cm}

{\bf Abstract}
\vspace{0.1cm}
\end{center}
\begin{quotation}
Recently, many works have tried to realize cosmological accelerated expansion in string theory models in the asymptotic regions of field space, with a typical scalar potential $V(\varphi)$ having an exponential fall-off $e^{-\gamma\, \varphi}$. Those attempts have been plagued by the fact that $V$ is too steep, namely $\gamma \geq 2/\sqrt{d-2}$ in a $d$-dimensional spacetime. We revisit the corresponding dynamical system for arbitrary $d$ and $\gamma$, and show that for an open universe ($k=-1$), there exists a new stable fixed point $P_1$ precisely if $\gamma > 2/\sqrt{d-2}$. Building on the recent work arXiv:2210.10813, we show in addition that cosmological solutions asymptoting to $P_1$ exhibit accelerated expansion in various fashions (semi-eternal, eternal, transient with parametrically controlled number of e-folds, or rollercoaster). We finally present realizations in string theory of these cosmological models with asymptotically accelerating solutions, for $d=4$ or $d=10$. We also show that these solutions do not admit a cosmological event horizon, and discuss the possibility of this being a generic feature of quantum gravity.
\end{quotation}

\end{titlepage}

\tableofcontents

\section{Introduction and summary}

Recent years have witnessed a renewed interest in string theory for the construction of viable cosmological models. A lot of effort has been put into constructing and testing de Sitter solutions, namely solutions from string theory that exhibit a 4-dimensional (4d) spacetime with a positive cosmological constant. Those could reproduce the observed accelerated expansion of our current universe, or serve as a first approximation for an early universe phase of inflation. The outcome of these attempts is that well-controlled examples of de Sitter solutions are difficult to find, if they exist at all (see e.g.~\cite[Sec. 5]{Junghans:2023lpo} for a recent account). The focus has then shifted recently to the study of rolling scalar fields $\{ \varphi^i \}$ in a positive scalar potential $V$, in $d$-dimensional gravitational models of the form
\beq
{\cal S} = \int \d^d x \sqrt{|g_d|} \left(\frac{M_p^{d-2}}{2}\, {\cal R}_d - \frac{1}{2} g_{ij}\, \del_{\mu} \varphi^i\, \del^{\mu} \varphi^j - V(\varphi^k)  \right) \ ,
\eeq
where $M_p$ is the Planck mass. Phenomenologically, such models in $d=4$ are known to provide valid cosmological models with accelerated expansion, namely inflation or quintessence models. The question is then whether those can be obtained from string theory.

Since a de Sitter solution would correspond to a critical point of $V$, a natural guess is that a potential with a small slope would also provide an observationally valid model; while a correct intuition, we will see that this is not the only option. Characterising the ratio $|\nabla V|/V$, with slope $|\nabla V|=\sqrt{g^{ij} \del_i V \del_j V}$, that can be obtained in string theory models, has thus been an important activity leading to the de Sitter conjectures of the swampland program (see e.g.~\cite{Agmon:2022thq} for a recent review). A particular focus has been given to the asymptotics of field space: those typically correspond to regions where string corrections are under control (e.g.~small string coupling, or large volumes corresponding to negligible higher derivative corrections). In the asymptotics, the following characterisation has been proposed: string theory potentials have been conjectured \cite{Bedroya:2019snp, Rudelius:2021oaz} to obey
\beq
\frac{|\nabla V|}{V} \geq \frac{2}{\sqrt{d-2}} \ , \label{ratiointro}
\eeq
with $M_p=1$, in a $d$-dimensional spacetime. This conjecture suffers no known counter-example.

String theory potentials typically have an exponential dependence on the (canonically normalized) fields. Considering as we will do only a single field, an exponential potential can be written as $V= V_0 \, e^{- \gamma\, \varphi}$, and the above is then nothing but a lower bound on the exponential rate $\gamma$. This lower bound on $\gamma$ seems naively too large for an observationally valid accelerated expansion model. Indeed, it is not compatible with single-field slow-roll inflation and related observational constraints \cite{Planck:2018jri}, and it is in tension with a valid single-field quintessence model \cite{Agrawal:2018own}; see also the recent extensive study \cite{Schoneberg:2023lun}. More dramatically, a dynamical system analysis \cite{Copeland:1997et, Tsujikawa:2013fta, Rudelius:2022gbz, Andriot:2022xjh, Shiu:2023nph, Shiu:2023rxt} shows that a stable fixed point in the far future, that we denote $P_2$, would only admit acceleration for
\beq
P_2\, ,\ \text{acceleration:}\qquad \gamma < \frac{2}{\sqrt{d-2}} \ . \label{P2intro}
\eeq
Note also that the requirement of stability, i.e.~$P_2$ being an attractor, corresponds to the same inequality. Comparing \eqref{ratiointro} and \eqref{P2intro} appears to ruin hopes of realizing accelerated expansion in string theory models, at least in the asymptotics of field space (and asymptotics in time), for which corrections are under control. This matter has been the topic of many recent works, including \cite{Russo:2018akp, Russo:2019fnk, Brinkmann:2022oxy, Rudelius:2022gbz, Andriot:2022xjh, Calderon-Infante:2022nxb, Marconnet:2022fmx, Shiu:2023nph, Shiu:2023rxt, Cremonini:2023suw, Hebecker:2023qke, Freigang:2023ogu, VanRiet:2023cca}.

However, upon closer examination of the situation, we see two important loopholes. First, having no acceleration at a future fixed point does not mean that acceleration cannot occur before. It can be realized during a transient phase (as is actually expected for inflation), as was first noticed in \cite{Townsend:2003fx}, but also it can be realized for an infinite amount of time, as was pointed out in \cite{Chen:2003dca, Andersson:2006du}, since the fixed point only corresponds to the asymptotic limit $t\rightarrow \infty$. Second, the previously mentioned dynamical system and fixed point analysis was carried out on a spatially flat spacetime. Using a Friedmann-Lema\^itre-Robertson-Walker (FLRW) metric, this corresponds to $k=0$. The analysis can be extended to $k=-1$, namely a spatially curved, open universe. This was done in \cite{Halliwell:1986ja} for $d=4$, and we will provide here a more complete analysis, with arbitrary $d$ and $\gamma$. With this extension, one finds a new stable fixed point $P_1$ leading to different physics. No accelerated expansion is occurring at $P_1$ itself, but cosmological solutions in its vicinity as well as away from $P_1$ exhibit accelerated expansion for an infinite amount of time, as described before. Those solutions are in addition able to accumulate an infinite number of e-folds while asymptoting to $P_1$. Other solutions also allow for transient acceleration, with a parametric control on the number of e-folds. Last but not least, as we will show in this paper, a necessary condition for $P_1$ with $k=-1$ to exist is
\beq
P_1\, ,\ \text{existence:}\qquad \gamma > \frac{2}{\sqrt{d-2}} \ . \label{P1intro}
\eeq
Note that $P_1$ is then automatically stable, i.e.~an attractor. Given the claim \eqref{ratiointro} and the possibility of having accelerating solutions around $P_1$, \eqref{P1intro} brings back the hope to realize asymptotic accelerated expansion in string theory, as was recently discussed in \cite{Marconnet:2022fmx}. Note that ``asymptotic'' does not refer here to the fixed point itself, but to the infinite time before it.

Realizing accelerating solutions in string theory is actually not a new idea. Following the time-dependent compactifications of \cite{Townsend:2003fx}, several solutions coming from string/M-theory, and  exhibiting accelerated expansion for a finite period of time, were subsequently constructed in \cite{Ohta:2003pu,Ohta:2003ie,Ohta:2004wk,Roy:2003nd,Gutperle:2003kc,Emparan:2003gg}. It was thus realized that transient acceleration is in fact generic in flux compactifications, see \cite{Townsend:2003qv} for a review. Until recently, however, all known examples of transient acceleration coming from string/M-theory were only able to produce a number of e-folds of order one, and were therefore thought to be unsuitable as models of inflation \cite{Chen:2003dca,Wohlfarth:2003kw}. This common lore was recently found to be false in \cite{Marconnet:2022fmx}, which showed that cosmologies exhibiting transient accelerated expansion with parametric control of e-folds are in fact generic in flux compactifications of 10d supergravity, provided one allows for an open universe with $k=-1$. In the present paper we review some models of \cite{Marconnet:2022fmx} and also provide a new string theory realization of an accelerated expanding universe. Whether or not such accelerating cosmological solutions are observationally valid remains to be seen; we come back to this matter in the Outlook.\\

We start this paper by carrying out in Section \ref{sec:general} a general dynamical system analysis of a single scalar field model, minimally coupled to gravity, in a $d$-dimensional FLRW spacetime, $d>2$, with an arbitrary scalar potential $V \geq0$. The spatial curvature, related to $k=0,\pm1$, as well as the ratio $\gamma=-V'/V$ at a given fixed point, are left free. We find the fixed points of the system, summarizing them in Table \ref{tab:sumfixedpoints}, study whether they allow for acceleration, and further analyse their stability. This analysis and the corresponding phase space are summarized graphically in Figure \ref{fig:PhaseSpace}.

We then restrict to $k=-1$ and to an exponential potential, with exponential rate $\gamma$, and study in Section \ref{sec:sol} the solutions that asymptote to the stable fixed point $P_1$. We first provide, with arbitrary $d, \gamma$, analytic expressions for classes of solutions in the vicinity of $P_1$, some exhibiting accelerated expansion. We then show numerically the complete solutions in phase space, and discuss the various ways this acceleration is realized. We also show that these solutions, asymptoting to $P_1$ in the future, have no cosmological event horizon; we add a few words on this observation below.

While all the above is a pure cosmological analysis, we finally show in Section \ref{sec:string} that these solutions can be realized within string theory models. We first review some results of \cite{Marconnet:2022fmx}: there the authors obtained appropriate $d=4$ two-scalar models with $k=-1$, as consistent truncations of 10d type IIA supergravity compactifications, with the 6d compact manifold being Calabi-Yau, Einstein, or Einstein-K\"{a}hler. These models always admit consistent sub-truncations to a single scalar with an exponential potential. Here, we also show that 10d massive IIA supergravity can itself provide a realization of such a model in $d=10$, where the scalar field is the dilaton and the exponential potential is generated by the Romans mass. Provided that these settings can be realized in the classical string regime, as argued for in Subsection \ref{sec:classical}, we conclude that cosmological solutions with asymptotic accelerated expansion can be realized in string theory.

The difficulty in string theory, and more generally quantum gravity, is therefore not to get asymptotic or eternal accelerated expansion, contrary to what the attempts with $k=0$, or with de Sitter solutions, suggest. From the present work, one could rather infer the following claim: {\sl in quantum gravity, the difficulty is to get solutions with cosmological event horizons.} This is in line with an absence of fully stable (instead of metastable) de Sitter solutions, and with the solutions in asymptotic accelerated expansion obtained here for $d\geq 3$. Interestingly, the notion of cosmological event horizon is intrinsically asymptotic (see e.g.~\eqref{de}); having a claim related to this concept then fits well with most swampland conjectures, that also tend to be asymptotic. Claiming the {\sl absence of cosmological event horizons in quantum gravity} is also in line with the absence of a holographic description for a cosmological setting, where a horizon at future infinity would have played the role of holographic boundary. The difficulty in getting such a holographic description is known for a de Sitter spacetime, related there to the non-unitarity of the would-be CFT. A similar conclusion was reached in \cite{Bedroya:2022tbh} after investigating power-law accelerating solutions. It would be interesting to investigate in more depth such a {\sl no cosmological horizon conjecture}.\\

Given that we find interesting cosmological solutions that require $k=-1$, one may wonder how natural this is in string theory; let us make a final comment on this matter. At first glance it seems that $k$ is an input that one can choose in any string theory construction. However, as formalised with the concept of the string landscape \cite{Bousso:2000xa, Susskind:2003kw}, it became apparent that string theory may have a huge number of vacua. It was then argued in \cite{Freivogel:2005vv} that in the early universe, transitions between those vacua would necessarily lead to $k=-1$. The reason for this is the tunneling between vacua as described by Coleman and de Luccia \cite{Coleman:1980aw}, which necessarily requires $k=-1$ after tunneling. This had led to the idea that $k=-1$ is not only possible but actually necessary in string theory. However, this is too strong a claim, as was shown in several more recent papers that find other $k$ values to be possible as well \cite{Buniy:2006ed, Horn:2017kmv, Cespedes:2020xpn, Cespedes:2023jdk}. Thus, we conclude that while not absolutely necessary, an external spacetime with $k=-1$ is still natural in string theory. We come back in the Outlook on how realistic related cosmological solutions can be.

\section{Setup and fixed point analysis}\label{sec:general}

In this section, we introduce the $d$-dimensional theory that will serve as our cosmological model. We then perform a dynamical system analysis of its equations, identifying its fixed points, their stability, and the regions with an accelerating universe. Our results are summarized in Table \ref{tab:sumfixedpoints} and Figure \ref{fig:PhaseSpace}.

\subsection{$d$-dimensional setup}

We consider a $d$-dimensional theory, $d>2$, describing a single scalar field minimally coupled to gravity
\beq\label{EFT}
{\cal S} = \int \d^d x \sqrt{|g_d|} \left(\frac{M_p^{d-2}}{2}\, {\cal R}_d - \frac{1}{2} \del_{\mu} \varphi\, \del^{\mu} \varphi - V(\varphi)  \right) \ ,
\eeq
where the field $\varphi$ is canonically normalized, and $V$ is the scalar potential. In the following, we set the Planck mass $M_p=1$. We focus on solutions with Friedmann-Lema\^itre-Robertson-Walker (FLRW) metric
\beq\label{eq:metric}
\d s^2 = - \d t^2 + a(t)^2 \left(\frac{\d r^2}{1 -k r^2} + r^2 \d \Omega^2 \right) \ , \quad k=0,\pm 1 \ , \ a(t)>0 \ ,
\eeq
with scale factor $a(t)$, and $k$ determining the curvature of the $(d-1)$-dimensional space. We also restrict ourselves to a homogeneous scalar field. Then, the equations of motion (e.o.m.) are given by the two Friedmann equations and the field e.o.m.,\footnote{A formal linear relation between the three equations can be found in e.g.~\cite[(A.2)]{Andriot:2022brg}. One consequence is that satisfying the first Friedmann equation together with the field e.o.m. implies that the second Friedmann equation is satisfied, as long as $H \neq 0$.} namely
\bea
& \frac{(d-1)(d-2)}{2} \left( H^2 + \frac{k}{a^2} \right) = \rho \ ,\nn\\
& (d-2) \frac{\ddot{a}}{a} + \frac{d-3}{d-1} \rho + p =0  \ \Leftrightarrow \ \dot{H} - \frac{k}{a^2} + \frac{\rho + p}{d-2} = 0 \ , \label{EOMs}\\
& \ddot{\varphi} + (d-1) H \dot{\varphi} + V' = 0 \ ,\nn
\eea
where the Hubble parameter, the energy density and the pressure are given by
\beq
H= \frac{\dot{a}}{a} \ ,\quad \rho = \frac{1}{2} \dot{\varphi}^2 + V \ ,\quad p = \frac{1}{2} \dot{\varphi}^2 - V \ .
\eeq
The dot stands for $\del_t$ and the prime for $\del_{\varphi}$.

In this context, it is useful to introduce the equation of state parameter $w =\frac{p}{\rho}$, for which we restrict ourselves to $V>0$. Using the second Friedmann equation, one shows that having an accelerating universe amounts to
\beq
\ddot{a} \geq 0 \ \Leftrightarrow \ w \leq -\frac{d-3}{d-1} \ ,\label{condacc}
\eeq
where we include for future convenience the case without acceleration, $\ddot{a}=0$.

\subsection{Dynamical system and fixed points}\label{sec:2.2}

We perform a dynamical system analysis on the previous equations. Similar analyses were carried-out for exponential potentials in $d=4$ \cite{Copeland:1997et, Tsujikawa:2013fta}, and arbitrary $d$ in \cite{Rudelius:2022gbz, Andriot:2022xjh}. A more thorough analysis was recently performed in \cite{Shiu:2023nph, Shiu:2023rxt} in a multifield situation. All these works focused on $k=0$; the main novelty here is that we allow for $k=0,\pm1$, and $k=-1$ will provide interesting new physics. We also allow for a generic positive scalar potential, the only quantity that will matter is the ratio $\gamma = -\left. V'/V \right|_0$ at a fixed point. This analysis has been performed in \cite{Marconnet:2022fmx} for $d=4$ and specific values for $\gamma$ coming from certain classes of string theory compactifications, a situation we will come back to; here we work with arbitrary $d, k$ and $\gamma$.

We introduce the following variables
\beq
N= \ln a \ ,\quad x = \frac{\dot{\varphi}}{H \sqrt{(d-1)(d-2)}} \ ,\quad y = \frac{\sqrt{2V}}{H \sqrt{(d-1)(d-2)}} \ ,\qquad H\neq 0 \ ,\ V>0 \ .\label{variables}
\eeq
Using the field e.o.m.~and the second Friedmann equation (without appearance of $k$), we obtain the following system
\bea
& \frac{\d x}{\d N} = - \frac{\sqrt{(d-1)(d-2)}}{2}\, \frac{V'}{V} \, y^2   - x\, \Big( d-2 - x^2 (d-2) + y^2 \Big) \ , \label{system}\\
& \frac{\d y}{\d N} = y \left( \frac{\sqrt{(d-1)(d-2)}}{2}\, \frac{V'}{V} \, x +1 +  x^2 (d-2) - y^2 \right)  \ .\nn
\eea
Because $k$ is not fixed here, we proceed differently than in previous references, and do not use the first Friedmann equation when deriving \eqref{system}; we will come back to that equation later.

We now look at the fixed points, given by
\beq
\frac{\d x}{\d N} = \frac{\d y}{\d N} = 0 \ ,
\eeq
and we introduce the convenient notations at a given fixed point
\beq
\gamma \equiv -\left. \frac{V'}{V}\right|_0 \ ,\qquad \lambda= \gamma\, \frac{\sqrt{(d-1)(d-2)}}{2} \ . \label{gammadef}
\eeq
To derive the system \eqref{system}, we considered the restriction $V>0$. It is however possible to extend the fixed point analysis to include the case $V=0$, i.e.~$y=0$. We get in that case the following fixed points
\bea
& P_0:\quad (x,y)=(0,0) \ ,\\
& P_{\pm}:\quad (x,y)=(\pm 1, 0)\ ,\nn
\eea
and we refer to \cite{Marconnet:2022fmx} for more details about them.

We turn to $y \neq 0$ and rewrite the fixed point equations as
\bea
\lambda \, y^2 & = x \left( d-1 - \lambda\, x \right) \ , \label{eq1}\\
 y^2 & =  x^2 (d-2) - \lambda\, x + 1  \ .\nn
\eea
If $\lambda =0$, we get $x=0$ and $y=\pm 1$. This fixed point will be included below in $P_2$. We then consider $\lambda \neq 0$. The equations above get rewritten into
\bea
&  x^2 \, \lambda(d-1) + \lambda  \, (-\lambda  x +1) = x\, (d-1) \ ,\label{eq2}\\
& (d-1)\, y^2 = 1 -  \frac{x}{\lambda} \Big(\lambda^2 -(d-1)(d-2) \Big) \ ,\nn
\eea
which are solved by the following fixed points
\bea
& P_1:\quad  (x,y)=\left(\frac{1}{\lambda},\, \pm \frac{\sqrt{d-2}}{\lambda} \right) = \left( \frac{2}{\gamma \sqrt{(d-1)(d-2)}},\, \pm \frac{2}{\gamma \sqrt{d-1}} \right) \\
& P_2:\quad  (x,y)=\left( \frac{\lambda}{d-1},\, \pm \sqrt{1 - \frac{\lambda^2}{(d-1)^2} } \right)  = \left( \frac{\gamma}{2}\sqrt{\frac{d-2}{d-1}} ,\, \pm \sqrt{1 - \frac{\gamma^2}{4} \frac{d-2}{d-1}} \right)
\eea
where $P_2$ exists iff $\lambda^2 < (d-1)^2$, i.e.~$\gamma^2 < 4\, \frac{d-1}{d-2} $ ($P_2$ is defined for $y\neq0$).

We finally turn to the first Friedmann equation: it gets rewritten as
\beq
x^2+y^2 =  1+ \frac{k}{\dot{a}^2} \ .\label{1stFried}
\eeq
Let us consider its impact on each fixed point. To start with, $P_{\pm}$ require $k=0$, while $P_0$ requires $k=-1$ together with the value at the fixed point $\dot{a}_0^2 = 1$. $P_2$ then requires $k=0$. This fixed point was found in previous references with $k=0$, which made use from the start of the resulting first Friedmann equation, contrary to here. Last but not least, $P_1$ requires the following relation
\beq
P_1: \quad \gamma^2 = \frac{4}{d-2} \left( 1+ \frac{k}{\dot{a}_0^2} \right)^{-1} \ , \label{lambda+}
\eeq
allowing for $k=\pm 1, 0$. For $k=0$, it fixes $\gamma= \pm \frac{2}{\sqrt{d-2}}$, coinciding then with $P_2$ for this value of $\gamma$.

Having determined the fixed points, we finally study the possibility of having acceleration there. Let us start with $P_0$ and $P_{\pm}$. As a cosmological solution, $P_0$ corresponds to a Milne universe, which admits a scale factor linear in $t$. Therefore, $\ddot{a}(t)=0$, i.e.~there is no acceleration at this fixed point. For $P_{\pm}$, since $\dot{\varphi}\neq0,\, V=0$, one obtains there $w=1$ and $\ddot{a}(t)<0$, i.e.~a decelerating universe. We now turn to $P_1, P_2$ for which $y\neq0$ and $\gamma$ is well-defined. By definition, one has $w= \frac{x^2-y^2}{x^2 +y^2}$, and the condition for acceleration is \eqref{condacc}. Using equations \eqref{eq1} and \eqref{eq2}, $w$ gets rewritten as
\beq
w= - 1 + \frac{2}{(d-1)}\ x \lambda = -1 + x \gamma\, \sqrt{\frac{d-2}{d-1}}   \ . \label{weq}
\eeq
The condition \eqref{condacc} for acceleration becomes $x \lambda \leq 1$, i.e.~$x \gamma \leq 2/\sqrt{(d-1)(d-2)} $: considered on each fixed point, we obtain
\bea
& P_1: \quad 1 \leq 1 \ ,\\
& P_2:\quad \gamma^2 \leq \frac{4}{d-2} \ .
\eea
The condition for $P_2$ is well-known from previous references. Interestingly, for $P_1$ there is no constraint on $\gamma$: the solution at this fixed point is simply never accelerating as the inequality is saturated, i.e.~$\ddot{a} = 0$. Note in particular the preservation of this property when $P_1$ and $P_2$ coincide with $\gamma= \pm \frac{2}{\sqrt{d-2}}$. We summarize the results in Table \ref{tab:sumfixedpoints}.

\begin{table}[H]
  \begin{center}
    \begin{tabular}{|c|c|c|c|}
    \hline
 & & & \\[-8pt]
Fixed point $(x,y)$ & Allowed $k$ & Existence constraint & Acceleration  \\[5pt]
\hhline{====}
 & & & \\[-6pt]
$P_0: \ \left(0,\, 0 \right)$ & $k= -1$ & $ \dot{a}_0^2 = 1 $ & no ($\ddot{a}=0$) \\[3pt]
 & & & \\[-8pt]
 \hline
 & & & \\[-6pt]
$P_{\pm}: \ \left(\pm 1,\, 0 \right)$ & $k= 0$  & - & no ($\ddot{a}<0$) \\[3pt]
 & & & \\[-8pt]
 \hline
 & & & \\[-8pt]
$P_1: \ \left(\frac{2}{\gamma \sqrt{(d-1)(d-2)}},\, \pm \frac{2}{\gamma \sqrt{d-1}}  \right)$ & $k=0, \pm 1$ & $ \gamma^2 = \frac{4}{d-2} \left( 1+ \frac{k}{\dot{a}_0^2} \right)^{-1} $ & no ($\ddot{a}=0$) \\[3pt]
 & & & \\[-8pt]
 \hline
 & & & \\[-8pt]
$P_2: \ \left(\frac{\gamma}{2}\sqrt{\frac{d-2}{d-1}},\, \pm \sqrt{1 - \frac{\gamma^2}{4} \frac{d-2}{d-1}} \right)$ & $k=0$ & $0 \leq \gamma^2 < 4\, \frac{d-1}{d-2}$ & iff $\gamma^2 < \frac{4}{d-2}$ \\[5pt]
    \hline
    \end{tabular}
    \caption{Fixed points $(x,y)=\left(\frac{\dot{\varphi}}{H \sqrt{(d-1)(d-2)}},\, \frac{\sqrt{2V}}{H \sqrt{(d-1)(d-2)}} \right)$ in flow of $N$-folds or time, with $H\neq 0,\, V\geq 0$, and their properties. We denote for $V>0$ the ratio at a fixed point $\gamma = - \left. \frac{V'}{V}\right|_0 $.}\label{tab:sumfixedpoints}
  \end{center}
\end{table}

We conclude that $k=\pm1$ allows for new fixed points w.r.t.~$k=0$. Of particular interest for string theory realizations (see Section \ref{sec:string}), $P_1$ with $k=-1$ admits
\begin{equation}
\gamma^2 > \frac{4}{d-2} \ . \label{gammabound}
\end{equation}
Even though the fixed point itself does not allow for acceleration, we will see in Section \ref{sec:sol} that a (semi)-eternally accelerating universe can be found arbitrarily close by. Prior to this, let us study the stability of these fixed points.

\subsection{Stability}

We now study the stability of the fixed points listed in Table \ref{tab:sumfixedpoints}. As a non-linear autonomous system, the stability is determined by looking at the sign of the real parts of the eigenvalues of the Jacobian of the system \eqref{system}, for each fixed point. A stable fixed point is obtained if these real parts are negative. The Jacobian to consider is obtained by acting with $\del_x$ or $\del_y$ on the right-hand side of each equation in the system \eqref{system}. It is then given by the following matrix
\beq
M= \left(\begin{array}{cc} (3 x^2-1) (d-2) - y^2  & 2y\,(\lambda- x)  \\
  y\, ( 2 x (d-2)-\lambda)  &  1-\lambda \, x + x^2(d-2) - 3 y^2  \end{array}\right) \ .
\eeq
In the following, we give its expression at each fixed point, together with its determinant and trace
\bea
& P_0:\quad M=\left(\begin{array}{cc} - (d-2)   & 0 \\
0  &  1  \end{array}\right) \ ,\\
& {\rm det} M=-(d-2) \ ,\ {\rm Tr} M = -(d-3) \ ,\nn\\[8pt]
& P_{\pm}:\quad M=\left(\begin{array}{cc} 2 (d-2)  & 0 \\
 0 &  d-1 \end{array}\right)  \ ,\\
& {\rm det} M= 2(d-1)(d-2) \ ,\ {\rm Tr} M = 3d-5  \ ,\nn\\[8pt]
& P_1:\quad M=\left(\begin{array}{cc} - (d-2) \left(1 - \frac{2}{\lambda^2} \right)   & 2y_1 \lambda \left( 1\,   - \frac{1}{\lambda^2} \right) \\
-\lambda y_1 \left( 1  - \frac{2(d-2)}{\lambda^2} \right)  &  -\frac{2(d-2)}{\lambda^2}  \end{array}\right) \ ,\label{MP1}\\
& {\rm det} M=2(d-2)\left(1- \frac{d-1}{\lambda^2} \right) = 2(d-2)\left(1- \frac{4}{\gamma^2 (d-2)} \right) \ ,\ {\rm Tr} M = -(d-2) \ ,\nn\\[8pt]
& P_2:\quad M=\left(\begin{array}{cc} 1-d + \lambda^2  \frac{3d-5}{(d-1)^2}  & 2 y_2 \lambda \frac{d-2}{d-1}  \\
 y_2 \lambda \frac{d-3 }{d-1}  &  -2 + \lambda^2  \frac{2}{(d-1)^2}) \end{array}\right) \ ,\\
 &  {\rm det} M= 2 \left(d-1 - \lambda^2 \right)\left(1 - \frac{ \lambda^2}{(d-1)^2} \right) = 2 (d-1) \left(1 - \frac{\gamma^2 (d-2)}{4} \right)\left(1 - \frac{\gamma^2}{4} \frac{d-2}{d-1} \right)\ ,\nn\\
 & {\rm Tr} M = -(d+1)+\frac{3}{d-1} \lambda^2 = -(d+1)+\frac{3(d-2)}{4} \gamma^2  \ .\nn
\eea
The sign of the real parts of the eigenvalues can be read from ${\rm Tr} M $ and ${\rm det} M$. Indeed, for a $2\times 2$ matrix $M$, the eigenvalues are
\beq
\frac{1}{2} \left( {\rm Tr} M \pm \sqrt{({\rm Tr} M)^2 - 4\, {\rm det} M} \right) \ .
\eeq
If ${\rm det} M<0$, then the square root is real, and it is bigger than $|{\rm Tr} M|$, so the eigenvalues are real and there is one positive and one negative eigenvalue. The fixed point is then unstable (a saddle point). If ${\rm det} M>0$, then either the square root is real but smaller than $|{\rm Tr} M|$, or it is purely imaginary. In either case, the real parts of the eigenvalues then have the sign of ${\rm Tr} M$, with corresponding stability behaviours. Finally, if ${\rm det} M=0$, (at least) one eigenvalue vanishes, and the stability is determined by the sign of the other eigenvalue, that of ${\rm Tr} M$.

We apply the above to the fixed points. We start with $P_0$ for which the situation is clear
\beq
P_0: \quad \text{unstable (saddle point)} \ .
\eeq
For $P_{\pm}$, the two eigenvalues are real and positive, giving
\beq
P_{\pm}: \quad \text{unstable (unstable node)} \ .
\eeq
We turn to $P_2$. There, we recall we must have $\gamma^2 < 4\, \frac{d-1}{d-2} $, so the sign of ${\rm det} M$ depends solely on the condition for acceleration, $\gamma^2 < \frac{4}{d-2} $. Provided the latter holds, we can show that ${\rm Tr} M<0$. We conclude:
\bea
P_2, \, \gamma^2 < \frac{4}{d-2} \ (\ddot{a}>0):&\quad \text{stable (stable node)} \ ,\nn\\
P_2, \, \gamma^2 = \frac{4}{d-2} \ (\ddot{a}=0):&\quad \text{Lyapunov stable (neutral node)} \ ,\\
P_2, \, \frac{4}{d-2} < \gamma^2 < 4\, \frac{d-1}{d-2} \ (\ddot{a}<0):&\quad \text{unstable (saddle point)} \ .\nn
\eea
For the case $\ddot{a}=0$, we have one vanishing and one negative eigenvalue, falling in the category of ``Lyapunov stable''. In addition, one verifies that
\beq
P_2:\quad  ({\rm Tr} M)^2 - 4\, {\rm det} M = \left(\frac{\gamma^2 (d-2)}{4} + d-3 \right)^2 \geq 0 \ ,
\eeq
so eigenvalues are always real. This is why we get a stable node for the accelerating case and not a stable spiral, and we get a neutral node for $\ddot{a}=0$ and not a neutral center.

We finally turn to $P_1$. The sign of ${\rm det} M$ is precisely that of $-k$, which makes obvious the stability since ${\rm Tr} M<0$. In addition, for ${\rm det} M\geq 0$, let us as above determine whether the eigenvalues are real or complex conjugate. This is determined by the sign of
\beq
P_1:\quad  ({\rm Tr} M)^2 - 4\, {\rm det} M = (d-2) \left( d-10 + \frac{32}{\gamma^2 (d-2)} \right)\ .\label{TMdetM+}
\eeq
This quantity is positive or zero for $d\geq 10$, or for
\beq
\gamma^2 \leq \gamma_s^2 \equiv \frac{32}{(d-2) (10 - d)} \ \ {\rm with}\ d<10 \ ,\label{gammas}
\eeq
giving then 2 real eigenvalues. We conclude
\bea
P_1, \, k=+1:&\quad \text{unstable (saddle point)}\nn\\
P_1, \, k=0:&\quad \text{Lyapunov stable (neutral node)}\\
P_1, \, k=-1,\, d\geq 10,\, {\rm or}\ d<10,\, \gamma^2 \leq \gamma_s^2:&\quad \text{stable (stable node)}\nn\\
P_1, \, k=-1,\, d<10,\, \gamma^2 > \gamma_s^2:&\quad \text{stable (stable spiral)}\nn
\eea

We end the discussion by highlighting the new stable fixed point,  $P_1$ with $k=-1$. For $d<10$, it is a stable node with
\beq
\frac{4}{d-2} < \gamma^2 \leq  \frac{32}{(d-2) (10 - d)} \ ,\label{gammarange}
\eeq
and for $d\geq10$ it is a stable node for any $\gamma^2 >\frac{4}{d-2}$. This is an interesting range for string theory constructions as we will see in Section \ref{sec:string}.

\subsection{Graphical summary}

Inspired by \cite{Marconnet:2022fmx} (see e.g.~Figure 8), we provide an illustration of the phase space and fixed points found above. The system depends on the two real variables $(x,y)$ which can be of either sign, so the phase space can first be represented by the $(x,y)$-plane. The dynamical system is made of the two differential equations \eqref{system}, and the first Friedmann equation \eqref{1stFried} which is not differential and thus appears as a constraint. Interestingly, the latter is given in terms of $x^2+y^2$, i.e.~the distance to the origin, and it separates the plane in 3 zones: the circle $x^2+y^2=1$ is the phase space for $k=0$, beyond it corresponds to $k=1$, and inside the circle corresponds to $k=-1$. Note that the fixed points $P_2$ lie only on the circle.

The dynamical system is symmetric under the following two independent changes of sign
\bea
& y \rightarrow -y \ ;\\
& x \rightarrow -x \ ,\ \frac{V'}{V} \rightarrow -\frac{V'}{V} \ ;
\eea
so in the following, when indicating the fixed points, we will restrict to the upper right quarter of the plane. The symmetry in $y$ indicates that for any expanding universe (upper half plane), there is another solution with a contracting one (lower half plane). The symmetry with $x$ and $\frac{V'}{V}$ is related to the freedom in the sign of the scalar field.

Further relevant regions in phase space are given by the acceleration condition \eqref{condacc}. Using $w= \frac{x^2-y^2}{x^2 +y^2}$, it gets rewritten as
\beq
{\rm Acceleration:}\quad |y| > |x| \, \sqrt{d-2} \ .
\eeq
The lines saturating this inequality correspond to the no acceleration region (where $P_1$ lies) and provide further distinct regions of the phase space.

Last but not least, the stability of the fixed points depends on their position. For $P_2$, it is directly related to the acceleration or deceleration. For $P_1$, in addition to $k$, the stability behaviour depends on the value of $\gamma$ for $k=-1$. The change happens at a certain distance from the origin given by
\beq
\sqrt{x_1^2+y_1^2}\bigg|_{\gamma=\gamma_s} = \sqrt{\frac{10-d}{8}} \ ,
\eeq
strictly smaller than 1 and greater than 0 as it should.

\begin{figure}[H]
\centering
\includegraphics[width=0.7\textwidth]{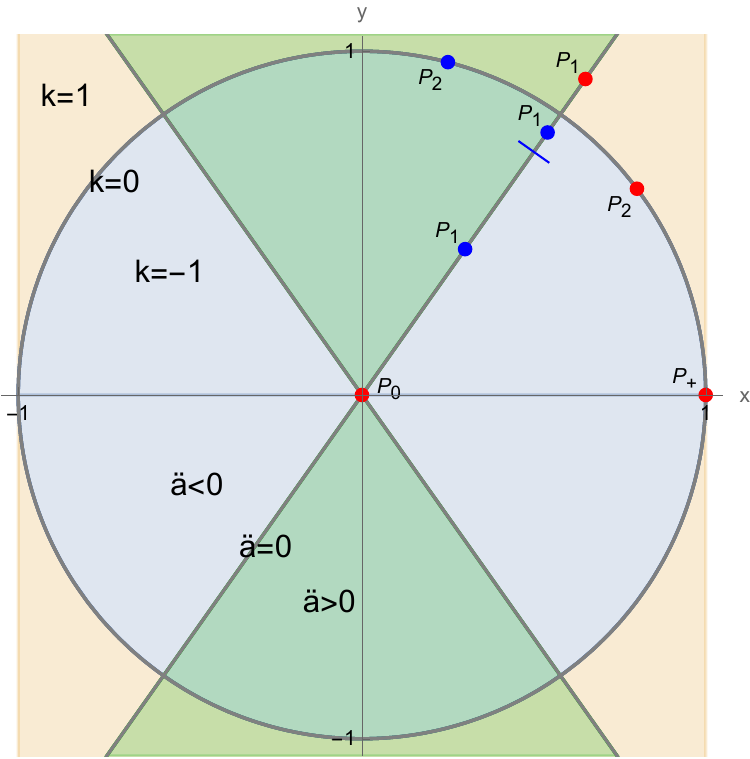}
\caption{Illustration of the phase space for the dynamical system considered, here in $d=4$. The physically relevant regions (value of $k$, acceleration) are indicated. Restricting to the upper right quarter of the plane, the fixed points $P_0, P_{+}, P_1, P_2$ are indicated in blue when stable and red when unstable. The short blue line separates the two different stabilities of $P_1$: the node (above) and the spiral (below). We refer to the main text for more details.}\label{fig:PhaseSpace}
\end{figure}

We illustrate the above in Figure \ref{fig:PhaseSpace}. Such illustration will be useful later to view cosmological solutions around the fixed points, as done already in \cite{Marconnet:2022fmx}.

\section{Cosmological solutions, acceleration and horizon}\label{sec:sol}

In this section, we determine cosmological solutions in the vicinity of the stable fixed point $P_1$ for $k=-1$, with a particular interest in those providing acceleration. We then investigate the way acceleration is realized and whether each solution admits a cosmological horizon.

We restrict to a positive exponential potential
\beq
V(\varphi)= V_0\, e^{-\gamma\, \varphi} \ , \qquad V_0, \gamma>0 \ , \label{V}
\eeq
where $\gamma= -V'/V$ is now a given constant everywhere in phase space. An extensive study of such cosmological solutions has been performed in \cite{Marconnet:2022fmx} for $d=4$ and particular values for $\gamma$ coming from certain classes of string theory compactifications; we will come back to the latter. We work here with a general $d\geq3$ and an arbitrary
\beq
\gamma > \frac{2}{\sqrt{d-2}} \ ,\label{gammaboundP1}
\eeq
where this lower bound is due to $k=-1$, see \eqref{gammabound}; we will see that this generically  leads to new interesting solutions.

Motivation for studying those solutions can be found in string theory models. In such models, it is a common situation that the scalar potential takes the form \eqref{V} in field space asymptotics, $\varphi \rightarrow \infty$. Focusing then on asymptotics (in field space but also in time), we are naturally interested in the physics close to attractive fixed points. As summarized in Figure \ref{fig:PhaseSpace}, those are $P_1, P_2$ for some values of $k,\gamma$. $P_2$ is attractive for $\gamma \leq \frac{2}{\sqrt{d-2}}$ (also corresponding to the accelerating regime). This fixed point was then studied in previous references. However, typical string theory constructions lead to $\gamma \geq \frac{2}{\sqrt{d-2}}$ in field and time asymptotics (see the strong de Sitter conjecture \cite{Bedroya:2019snp, Rudelius:2021oaz}), thus forbidding to reach such a $P_2$. On the contrary, $P_1$ allows for $\gamma$ in agreement with that conjecture as indicated in \eqref{gammaboundP1}, as long as $k=-1$.\footnote{$k=0$ would allow to saturate the conjecture's bound. It is however difficult to construct a string theory scalar potential precisely matching this bound.} In addition, it is then attractive, so this motivates us to focus on that fixed point. Being placed at the boundary of acceleration, solutions that asymptote to it can give accelerating universes, as we will see.

We thus turn to the study of solutions close to $P_1$ with $k=-1$. As seen in the stability analysis, one needs to distinguish
\bea
\text{Stable node:}&\quad \gamma \leq \gamma_s \nn\\
\text{Stable spiral:}&\quad \gamma > \gamma_s \label{P1around}\\
{\rm where}& \quad \gamma_s= \frac{4\sqrt{2}}{\sqrt{(10-d)(d-2)}} \ ,\nn
\eea
whenever restricting to $d<10$. We thus treat those two situations one after the other. For $d\geq 10$, the solutions found for the stable node with $\gamma \leq \gamma_s$ will be valid for any $\gamma$.

Before looking at the solutions close to $P_1$, let us add a word on the fixed point itself. In general, fixed points correspond by themselves to cosmological solutions. Phase space trajectories between fixed points are also solutions, and those asymptote to each fixed point solution in some late or early time limit. We therefore need to know the solution corresponding to $P_1$ with $k=-1$, in order to reproduce it in the asymptotics. It can easily be found by solving the equations \eqref{EOMs}, with the dynamical system variables $x,y$ being constant: we get
\bea
P_1,\ k=-1:\qquad & a(t) = \frac{\gamma \ (t-t_0)}{\sqrt{\gamma ^2-\frac{4}{d-2}}} \ ,\quad e^{\gamma\, \varphi(t)} = \frac{\gamma^2 V_0\ (t-t_0)^2}{2(d-2)}  \nn\\
\Leftrightarrow\ \ & a(t)=a_0\, (t-t_0) \ ,\quad \varphi(t)=\varphi_0 + \varphi_l\, \log (t-t_0) \ , \label{P1sol}\\
{\rm with} \ \ & a_0=\frac{\gamma }{\sqrt{\gamma ^2-\frac{4}{d-2}}} \, ,\ \varphi_0 = \frac{1}{\gamma}\log\left(\frac{\gamma^2 V_0}{2(d-2)}\right) \, ,\ \varphi_l = \frac{2}{\gamma} \ .\nn
\eea
This is a Milne universe with angular defect \cite{Marconnet:2022fmx}. The free constant $t_0$ can be absorbed into a redefinition of $t$. In the following, we will find solutions around $P_1$ which asymptote to \eqref{P1sol} in the limit $t\rightarrow \infty$; for this reason we also drop $t_0$ from now on. We now turn to these neighbouring solutions.

\subsection{Solutions asymptoting to the stable node}\label{sec:solnode}

To find solutions that approach the fixed point solution \eqref{P1sol} for $t \to \infty$, we make the following Ansatz with $p>0$
\begin{align}\label{eq:ansatz}
    a(t) &= a_0\, t \left(1+ \frac{a_1}{t^{p}}+ \frac{a_2}{ t^{2 p}}+ \frac{a_3}{t^{3 p}}+ \ldots\right)\ , \cr
    \varphi(t) &= \varphi_0 + \varphi_l \log(t) + \frac{\varphi_1} {t^{p}}+ \frac{\varphi_2} {t^{2 p}} + \frac{\varphi_3}{t^{3 p}} + \ldots \ .
\end{align}
The fixed point corresponds to the leading order terms, given by $a_0,\, \varphi_0, \, \varphi_l$ in \eqref{P1sol}. We then solve equations at first subleading order. Since there are only two independent equations, we only fix the two following free parameters
\beq
    p^\pm =\frac{d-2}{2} \pm \frac{2\sqrt{2}}{\gamma} \sqrt{1+ \gamma^2 \frac{(d-10) (d-2)}{32}}\ ,\quad \varphi_1^\pm =\frac{d-1}{4}\, a_1 \gamma\ p^{\pm}  \ ,\label{eq:p}
\eeq
meaning we find two solution families labeled with $\pm$, and parameterized by $a_1$. These solutions are valid for any $d\geq3$. Restricting to $d<10$, we can rewrite $p^{\pm}$ in terms of $\sqrt{1-\frac{\gamma^2}{\gamma^2_s}}$, requiring as expected $\gamma \leq \gamma_s$. In the special case $\gamma=\gamma_s$, the two families become only one.

As it should in the Ansatz, we verify that $p^{\pm}>0 \Leftrightarrow \gamma > 2/\sqrt{d-2}$, which is true here. We can proceed further and solve the equations order by order, with the resulting expressions for the coefficients becoming more complicated. While solving, we observe however a problem for $p=1$: in that case, all higher order terms vanish. In other words, the above solution does not exist for $p=1$. Before focusing on this matter, let us first consider $p^{\pm}\neq1$ and study the properties of our solutions.

\begin{figure}[H]
\centering
\includegraphics[width=0.7\textwidth]{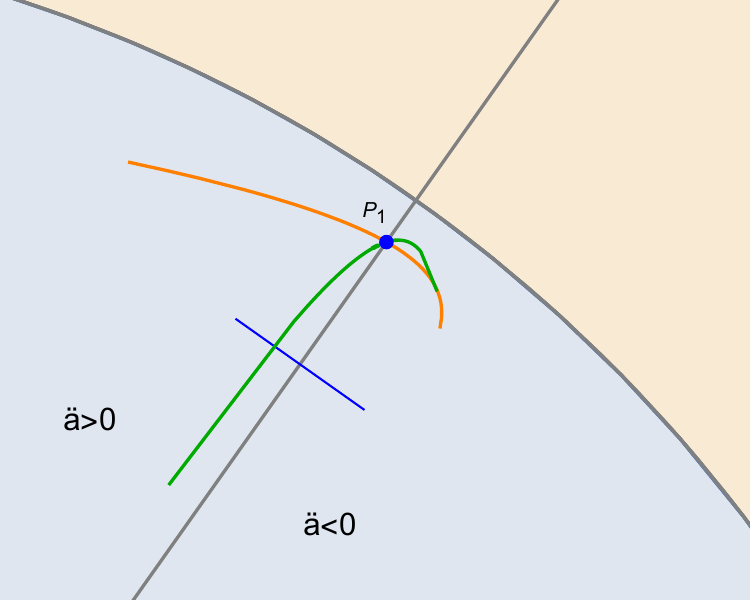}
\caption{Evolution of the dynamical variables $(x,y)$, defined in \eqref{variables}, for the solution families $+$ (orange) and $-$ (green). The solutions are given at first order by \eqref{eq:p}, here with $d=4,\gamma=\sqrt{2}+0.05$. The figure is a zoomed in part of Figure \ref{fig:PhaseSpace} around the stable node $P_1$ (blue point). The solutions are depicted for $a_1=\pm 1$, allowing us to verify that each branch is accelerating or decelerating.}\label{fig:P1sol1}
\end{figure}

In the large $t$ limit, properties can be read from the leading and subleading terms above. Of particular interest is the question of acceleration. From \eqref{eq:ansatz}, one has
\begin{equation}\label{eq:accel}
    \ddot{a}(t) = a_0 a_1 (p-1) p \ \frac{1}{t^{1+p}} + \mathcal{O}\left(t^{-(1+2p)}\right) \ .
\end{equation}
Since $a_0 p > 0$, the sign of $\ddot{a}(t)$ for our solutions is fixed by $a_1(p-1)$. Since we consider $p^\pm\neq1$, each family ($+$ or $-$) of solutions then has an accelerating and a decelerating branch, depending on the sign of $a_1$. In between, $\ddot{a}(t) =0$ when $a_1=0$, which corresponds to the fixed point $P_1$. We illustrate the two solution families and their branches in Figure \ref{fig:P1sol1}. Note that the two branches of one family are actually not continuously connected in terms of time, since they both asymptote to $P_1$ for $t\rightarrow \infty$.

As a technical remark, we indicate that the tangent vectors to the solution curves in Figure \ref{fig:P1sol1}, at the point $P_1$, are known analytically. Those correspond to the eigenvectors of the Jacobian \eqref{MP1}. They agree with the solution above.\\

We now come back to the case of $p=1$, where the previous solutions disappear. We verify the following
\bea
& d>4:\quad p^+>1 \ ,\ p^-=1 \Leftrightarrow \gamma = \gamma_m \equiv \frac{2 \sqrt{2}}{\sqrt{d-1}} < \gamma_s \ ,\nn\\
& d=4:\quad p^{\pm}=1 \Leftrightarrow \gamma=\gamma_s=\gamma_m \ ,\\
& d=3:\quad p^{\pm}<1 \ .\nn
\eea
It is therefore only for $d\geq4$ and for a specific $\gamma=\gamma_m$ that one can have $p=1$. Note the specificity of $d=4$, for which both solution families disappear when $\gamma=\gamma_m$, corresponding in addition to the upper bound $\gamma_s$! And it is actually for this very particular value that cosmological solutions around $P_1$ (stable node) were found from string theory compactifications in \cite{Marconnet:2022fmx}. The results of that reference will thus be helpful as we now explain.

We now consider the following Ansatz, which goes beyond \eqref{eq:ansatz} for $p=1$. This Ansatz is inspired from the explicit solutions of \cite{Marconnet:2022fmx} in $d=4$. There, solutions are first expressed in terms of $N$, the flow variable in the dynamical system. Relating it to $t$ thanks to $a(t)$, one can reach the following Ansatz
\bea
 \hspace{-0.2in} a(t) &= a_0\, t + a_l \log(t) +a_c +\frac{1}{t}\Big( a_{11} \log(t)+ a_{10} \Big) + \frac{1}{t^2} \Big( a_{22} (\log(t))^2 + a_{21} \log(t)+ a_{20} \Big) +\ldots\ , \nn\\
 \hspace{-0.2in} \varphi(t) &= \varphi_0 + \varphi_l \log(t) + \frac{1}{t} \Big(\varphi_{11} \log(t)+ \varphi_{10} \Big)+ \frac{1}{t^2} \Big( \varphi_{22} (\log(t))^2+ \varphi_{21} \log(t)+ \varphi_{20} \Big) +\ldots \ .\label{eq:ansatz2}
\eea
Solving the equations at the first subleading order, we find the following solution family (in terms of the parameters $a_l,a_c$)
\begin{align}
    \gamma &=\gamma_m=\frac{2 \sqrt{2}}{\sqrt{d-1}}\ , \nn\\
    \varphi_{11} &=\frac{1}{2}\, a_l \, \sqrt{\frac{(d-3)(d-1)}{d-2}}\ , \label{sol2}\\
    \varphi_{10} &= \frac{1}{2} \left( a_l\, \frac{5-2 d}{d-1} + a_c \right) \sqrt{\frac{(d-3)(d-1)}{d-2}} \ .\nn
\end{align}
The solution family requires to fix the value of $\gamma$ to $\gamma_m$. It can therefore be considered as replacing the previously lost solutions for $p=1$, and it does so by including logarithms in the Ansatz. Note though that for $d=4$ and $\gamma=\gamma_m=\gamma_s$, the overall number of solution family is only one, while all other cases admit two.

\begin{figure}[H]
\centering
\includegraphics[width=0.7\textwidth]{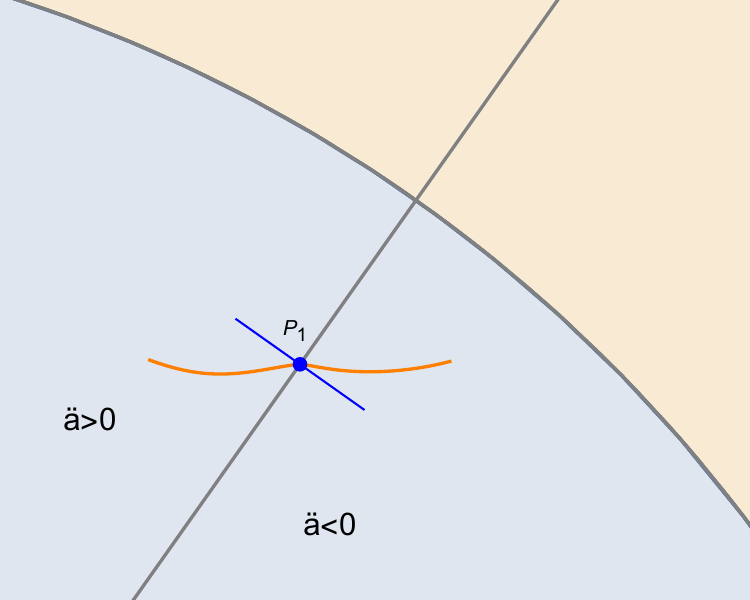}
\caption{Evolution of the dynamical variables $(x,y)$, defined in \eqref{variables}, for the solution family (in orange) given at first order by \eqref{sol2}, here with $d=4,\gamma=\gamma_m$. The figure is a zoomed in part of Figure \ref{fig:PhaseSpace} around the stable node $P_1$ (blue point). The solutions are depicted for $a_l=\pm 1, a_c=0$, allowing us to verify that each branch is accelerating or decelerating. We also observe the particularity of $d=4$ for which $\gamma_m=\gamma_s$, thus placing $P_1$ on the blue line; beyond this line, the stable node turns into a stable spiral.}\label{fig:P1sol2}
\end{figure}

Equations can be solved order by order, and the parameter $a_l$ remains free. As before, it determines the acceleration/deceleration for large $t$, since we have
\begin{equation}
    \ddot{a}(t) = -\frac{a_l}{t^2} + \mathcal{O}\left( \frac{\log(t)}{t^3}\right)\ .
\end{equation}
The solution family has two branches, one  accelerating and one decelerating, depending on the sign of $a_l$. For $a_l=0$, we recover $P_1$ with $\ddot{a}(t)=0$. We illustrate this solution family in Figure \ref{fig:P1sol2}.

As a final technical remark, we rewrite the Jacobian \eqref{MP1} at $P_1$ as follows
\beq
M_{\pm}  = \left(\begin{array}{cc} \frac{\gamma_m^2}{\gamma^2} -(d-2)   & \pm \frac{2}{\sqrt{d-2}} \left( d-2   - \frac{\gamma_m^2}{2 \gamma^2} \right) \\
\pm \sqrt{d-2} \left( \frac{\gamma_m^2}{\gamma^2} -1 \right)  &  -\frac{\gamma_m^2}{\gamma^2 }  \end{array}\right) \ ,
\eeq
where the sign refers to the sign in $y_1$. We recall that the eigenvectors of this matrix give the tangent directions to the solutions, at $P_1$. For $\gamma=\gamma_m$, the matrix becomes upper triangular. As a consequence, $(1,0)$ is an eigenvector. In other words, the tangent to the solution at $P_1$ is horizontal, which agrees with the orange line in Figure \ref{fig:P1sol2}.

\subsection{Solutions asymptoting to the stable spiral}\label{sec:solspiral}

We now look for solutions approaching the fixed point solution \eqref{P1sol} for $t \to \infty$, in the case $d<10$, $\gamma> \gamma_s$, corresponding to a spiral \eqref{P1around}. Such solutions were found in \cite{Marconnet:2022fmx} in $d=4$ for large $\gamma$. We generalize these solutions here to $3\leq d <10$, adapting appropriately the Ansatz to the following one
\begin{align}\label{eq:ansatz3}
    a(t) &= a_0 t \left( 1 + \frac{1}{t^{\frac{d-2}{2}}} \Big( a_{c1}\, \cos\big[ q\log (t)\big] + a_{s1}\, \sin\big[ q\log (t)\big] \Big) \right.\\
    &\qquad\qquad\ \left.+ \frac{1}{t^{d-2}} \Big(a_{c21}\, \cos\big[ q\log(t)\big]+a_{c22}\, \cos\big[ q\log (t)\big]^2 \right.\cr
    &\qquad\qquad\qquad\quad \, \left. + a_{s21}\, \sin\big[ q\log (t)\big]+ a_{s22}\, \sin\big[ q\log (t)\big]^2 \right.\cr
    &\qquad\qquad\qquad\quad \, \left.+ a_{cs}\, \cos\big[ q\log (t)\big]\sin\big[ q\log (t) \big] \Big)\ +\ \ldots \ \right)\ , \cr
    \varphi(t) &= \varphi_0 + \varphi_l \log(t) + \frac{1}{t^{\frac{d-2}{2}}} \Big( \varphi_{c1}\, \cos\big[ q\log (t)\big] + \varphi_{s1}\, \sin\big[ q\log (t)\big] \Big) \cr
    &\qquad\qquad\qquad\quad \ + \frac{1}{t^{d-2}} \Big(\varphi_{c21}\, \cos\big[ q\log (t)\big]+\varphi_{c22}\, \cos\big[ q\log (t)\big]^2 \cr
    &\qquad\qquad\qquad\qquad\qquad \, + \varphi_{s21}\, \sin\big[ q\log (t)\big]+ \varphi_{s22}\, \sin\big[ q\log (t)\big]^2\cr
    &\qquad\qquad\qquad\qquad\qquad \, + \varphi_{cs}\, \cos\big[ q\log (t)\big]\sin\big[ q\log (t)\big]\Big) +\ldots \ .\nonumber
\end{align}
The leading order is given by the $P_1$ solution \eqref{P1sol}. At the first subleading order we find
\begin{align}
q &= \frac{2\sqrt{2}}{ \gamma } \sqrt{\frac{\gamma^2}{\gamma_s^2} - 1}\ , \cr
\varphi_{c1} &= \frac{d-1}{8}\, \gamma \left(a_{c1}\, (d-2) - a_{s1}\, 2 q\right)\ , \label{P1sol3}\\
\varphi_{s1} &= \frac{d-1}{8}\, \gamma \left( a_{s1}\, (d-2) + a_{c1}\, 2 q \right) \ .\nn
\end{align}
We have a solution family depending on two real parameters: $a_{c1}, a_{s1}$.\footnote{One could consider introducing in the Ansatz \eqref{eq:ansatz3} a phase in the cosine and sine functions. Such a phase can however be absorbed in a redefinition of the coefficients $a_{c1}, a_{s1},$etc. Such a phase would then be a redundant parameter.} We also found a solution with $q\rightarrow -q$, but that change of sign can be compensated by $a_{s1} \rightarrow - a_{s1}$, with similar changes at higher orders, so we restrict ourselves to $q>0$.

\begin{figure}[H]
\centering
\includegraphics[width=0.7\textwidth]{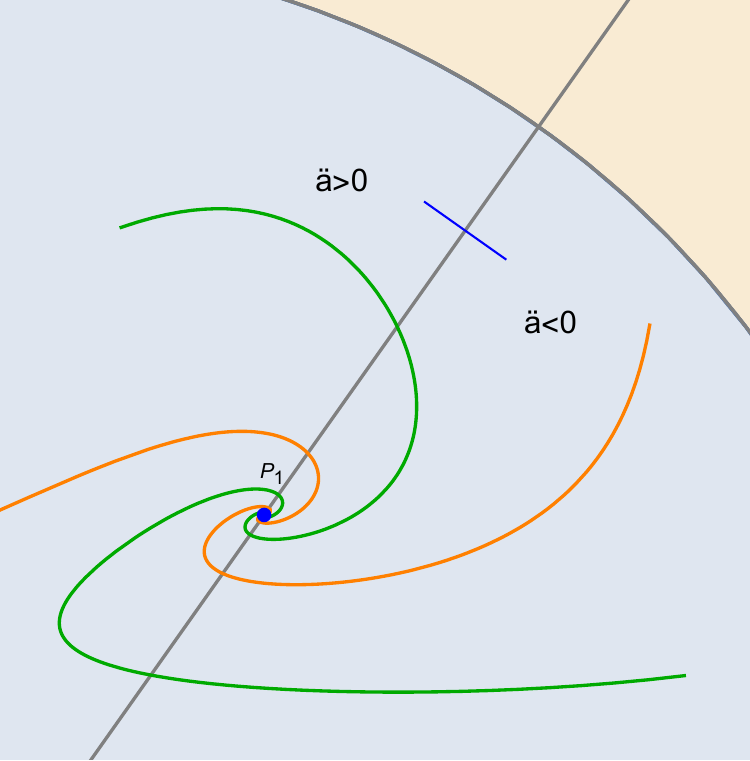}
\caption{Evolution of the dynamical variables $(x,y)$, defined in \eqref{variables}, for the solution family given at first order by \eqref{P1sol3}, here with $d=4,\gamma=\gamma_s+1.2$. The figure is a zoomed in part of Figure \ref{fig:PhaseSpace} around the stable node $P_1$ (blue point). We first verify that $P_1$ is beyond the blue line and is approached by spirals. The solutions are depicted for $(a_{c1},a_{s1})=\pm(1,1)$ (orange) and $\pm(1,-1)$ (green). This allows us to verify each branch of the same color is accelerating or decelerating.}\label{fig:P1sol3}
\end{figure}

We now turn to the question of acceleration. We compute at first subleading order
\begin{align}
    \ddot{a}(t) = \frac{\gamma}{4 \sqrt{\gamma ^2-\frac{4}{d-2}}}\ \frac{1}{t^{\frac{d}{2}}} &\Bigg( \Big(a_{c1}\, 4 (d-3) q+a_{s1} \left((d-6) d +8 -4 q^2\right)\Big) \sin\big[q \log (t)\big]  \\
    &\!\! -\Big(a_{s1}\, 4 (d-3) q -a_{c1} \left((d-6) d + 8 -4 q^2 \right)\Big) \cos\big[ q \log (t)\big]\Bigg)\ .\nonumber
\end{align}
For any given time $t$, if one solution is accelerating, flipping simultaneously the signs of $a_{c1}, a_{s1}$ gives a decelerating solution. We then have four branches of solutions depending on the signs of $a_{c1}, a_{s1}$, as illustrated in Figure \ref{fig:P1sol3}.

Whether a solution is accelerating or decelerating is time dependent. More precisely, for any given fixed parameters $a_{s1}$ and $a_{c1}$ the above solutions undergo an infinite number of accelerating and decelerating phases. This is easiest to see if $a_{s1}$ and $a_{c1}$ are chosen such that for example the prefactor of the cosine function vanishes. Then $\ddot{a}(t) \propto \sin\big[q\log(t)\big]$ changes its sign infinitely many times for $t \to \infty$. The same holds with both prefactors being non-zero.

\subsection{Horizon and acceleration}

As a cosmological solution, the stable fixed point $P_1$ with $k=-1$ is given in \eqref{P1sol}. In Section \ref{sec:solnode} and \ref{sec:solspiral}, we have determined cosmological solutions approaching the latter for $t \rightarrow \infty$. While $P_1$ is not a universe undergoing accelerated expansion, several of the solutions asymptoting to it are actually accelerating for an infinitely long time! Those can be completed numerically into full solutions, asymptoting in the past to another fixed point. Their acceleration phases are then realized in various manners that we describe below. Prior to this, let us first discuss an important common property to all these solutions, which is, the absence of an event horizon.

\subsubsection{No event horizon}

Given a cosmological solution with an $a(t)$, it is interesting to determine whether it admits an event horizon at a time $t_i$. To that end, one should compute the distance $d_e$ traveled by light until the end of time, denoted by $t_f$
\begin{equation}
d_e = a_i \int_{t_i}^{t_f} \frac{dt}{a(t)}\ ,
\end{equation}
where $a_i=a(t_i)$ and $t_f$ is the largest possible time; we take here $t_f=\infty$. If $d_e$ is finite, then the universe admits an event horizon of size $d_e$. This is the case for the de Sitter solution with $a(t)=a_i\, e^{H (t-t_i)}$ and a constant $H$, or for an accelerated expansion with $a(t)=a_i \, (t/t_i)^p$ and $p>1$. The boundary case $p=1$ without acceleration has no horizon, or equivalently has an infinite horizon size.

Let us now evaluate the horizon size $d_e$ for the cosmological solutions considered previously, that asymptote to $P_1$. The horizon is evaluated at a finite time $t_i$. We do not take $t_i$ to correspond to another asymptotic time, where the solution would match another fixed point or hit a singularity. Then, we can cut the integral in two pieces, the first one being between two finite times $t_i$ and $t_a> t_i$: the integral between those two is finite. We choose $t_a$ sufficiently large, in such a way that the solution between $t_a$ and $\infty$ can be approximated by an expansion around $t\rightarrow \infty$. More precisely, one verifies for the solution Ansatz \eqref{eq:ansatz}, \eqref{eq:ansatz2} and \eqref{eq:ansatz3} that at a sufficiently large $t$, one has
\beq
a(t)=a_0\ t \left( 1 + {\rm corr.} \right) \ , \ \quad |{\rm corr.}| < 1 \ .
\eeq
Picking $t_a$ in this way, we get
\bea
d_e & = a_i \int_{t_i}^{\infty} \frac{dt}{a(t)} = a_i \int_{t_i}^{t_a} \frac{dt}{a(t)} + a_i \int_{t_a}^{\infty} \frac{dt}{a(t)} \label{de}\\
& = ({\rm finite}) + \frac{a_i}{a_0} \int_{t_a}^{\infty} \frac{dt}{t \left( 1 + {\rm corr.} \right)} >  ({\rm finite}) + \frac{a_i}{a_0} \int_{t_a}^{\infty} \frac12 \frac{dt}{t} = \infty \ . \nn
\eea
We conclude there is no event horizon for the solutions asymptoting to $P_1$, i.e.~the horizon size is infinite. This is true despite the fact that some solutions are accelerating at any finite time. Thus, we conclude that eternal accelerated expansion does not necessarily lead to horizons. In Section \ref{sec:string} below, we will discuss that eternally accelerating cosmologies do arise in quantum gravity. The difficulty in finding de Sitter solutions could then be attributed to the existence of a finite event horizon instead of eternal accelerated expansion.

\subsubsection{Stable node: semi-eternal or eternal acceleration; transient acceleration with parametric control of e-folds}

\begin{figure}[H]
\centering
\includegraphics[width=0.7\textwidth]{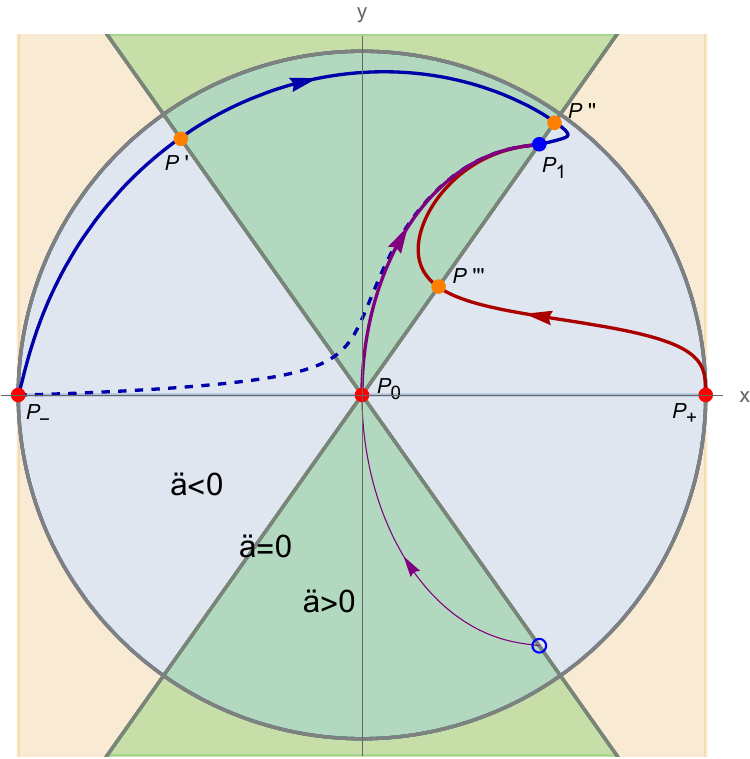}
\caption{Complete cosmological solutions asymptoting to $P_1$ as a stable node, displayed as trajectories in the phase space of Figure \ref{fig:PhaseSpace}, here with $d=4$ and $\gamma=\sqrt{\frac{8}{3}} - 0.05$. The purple solution exhibits eternal acceleration, the red ones semi-eternal acceleration and the blue ones transient acceleration. The plain versus dashed blue curves correspond to solutions with different numbers of e-folds, on which we have parametric control. The thin purple curve corresponds to the geodesic completion of the thicker one. We refer to the main text for more details.}\label{fig:solsP1node}
\end{figure}

In Subsection \ref{sec:solnode}, we have obtained various solution families asymptoting to $P_1$ when the latter is a stable node. Each solution family was found to have locally an accelerating and a decelerating branch, depending on which side of the acceleration cone they approached $P_1$. This was illustrated in Figure \ref{fig:P1sol1} and \ref{fig:P1sol2}. Numerically, one can solve the system of equations and find complete solutions evolving in phase space. Those necessarily asymptote in the past to one of the other fixed points appearing in Figure \ref{fig:PhaseSpace}. We display in Figure \ref{fig:solsP1node} these complete solutions. As we will see, they can have transient or (semi)-eternal acceleration phases. Most of our discussion is drawn from the results of \cite{Marconnet:2022fmx}, even though the solutions considered here are more general.

Let us start with a specific solution which asymptotes to $P_1$ (at $t\rightarrow \infty$) and asymptotes in the past to $P_0$ (at $t\rightarrow 0$). This phase space trajectory never leaves the acceleration cone, thus exhibiting eternal acceleration. While the solution at $P_0$ is a regular Milne universe ($a(t) \sim t$), the solution in its vicinity ($a(t)$ corrected at $\mathcal{O}(t^3)$) is nothing but a de Sitter universe. Indeed, it was shown in \cite{Marconnet:2022fmx} that, up to and including terms of order $\mathcal{O}(t^4)$, the spacetime metric becomes that of de Sitter space in the vicinity of $P_0$
\eq{\label{desitter}
\d s^2_{dS}=-\d t^2+l^2 \sinh^2\left(\frac{t}{l}\right)\, \left(\d \chi^2 + \sinh^2(\chi) \d \Omega^2 \right)
~,}
where the radius $l$ is related to the scalar curvature ${\cal R}_4$ of de Sitter via $l^2=12/{\cal R}_4$. Note that this is the de Sitter metric in hyperbolic slicing (with $k=-1$), different from the more familiar one with an exponential scale factor and $k=0$. Recall this is not an {\it asymptotic} de Sitter, since $t=0$, where the spacetime becomes a regular Milne universe, is reached at finite proper time in the past. One can further show \cite{Marconnet:2022fmx} that the flow parameter $N$, corresponding to the number of e-folds, is related to time via
\eq{\label{ps298}
\frac{t}{l} \sim e^N \left[ 1+\mathcal{O}\left( e^{2N}\right)\right]
~,}
so that $t=0$ corresponds to $N\rightarrow-\infty$. This allows one to conclude that the number of e-folds accumulated between $t=0$ and any finite $t>0$ is infinite. Likewise, we can conclude that asymptoting to $P_1$ leads to an infinite number of e-folds since $t \sim e^N$ and thus for $t\to \infty$ we accumulate an infinite number of e-folds.

Finally, let us point-out that the solution between $P_1$ and $P_0$ is actually not geodesically complete. It can be completed beyond the point $P_0$ in the past to $t<0$, by gluing together its mirror trajectory in the lower half of the phase space disk. The scale factor vanishes as $t\rightarrow0$, but it can be shown that the total 4d spacetime remains smooth in this limit. The phase trajectory actually goes through $P_0$, instead of asymptoting to it. We refer to \cite{Marconnet:2022fmx} for more details.\\

Having understood that the solution between $P_0$ and $P_1$ allows for eternal acceleration and an infinite number of e-folds, it is easier to understand the features of the other solutions. As displayed in Figure \ref{fig:solsP1node} there are two possible types of solutions: those asymptoting to $P_1$ accelerating (and asymptoting in the past to $P_+$), or those asymptoting to $P_1$ decelerating (and asymptoting in the past to $P_-$). One such solution of the former type starts accelerating at the point $P'''$, when entering the acceleration cone, and from there is eternally accelerating since $P_1$ is only an asymptotic point; we then qualify such solutions as having semi-eternal acceleration and leading to an infinite number of e-folds. The other solutions accelerate between two points only, $P'$ and $P''$, and thus have a transient acceleration. Interestingly though, the point $P'$ can be brought as close as desired to $P_0$, with the result of increasing the number of e-folds. This was shown numerically in \cite{Marconnet:2022fmx} and can be understood by the fact that the phase space trajectory becomes close to the one with eternal acceleration, which has an infinite number of e-folds coming from the vicinity of $P_0$. For all these solutions, we then have a parametric control of the number of e-folds.

\subsubsection{Stable spiral: rollercoaster cosmology; transient acceleration with parametric control of e-folds}

\begin{figure}[H]
\centering
\includegraphics[width=0.7\textwidth]{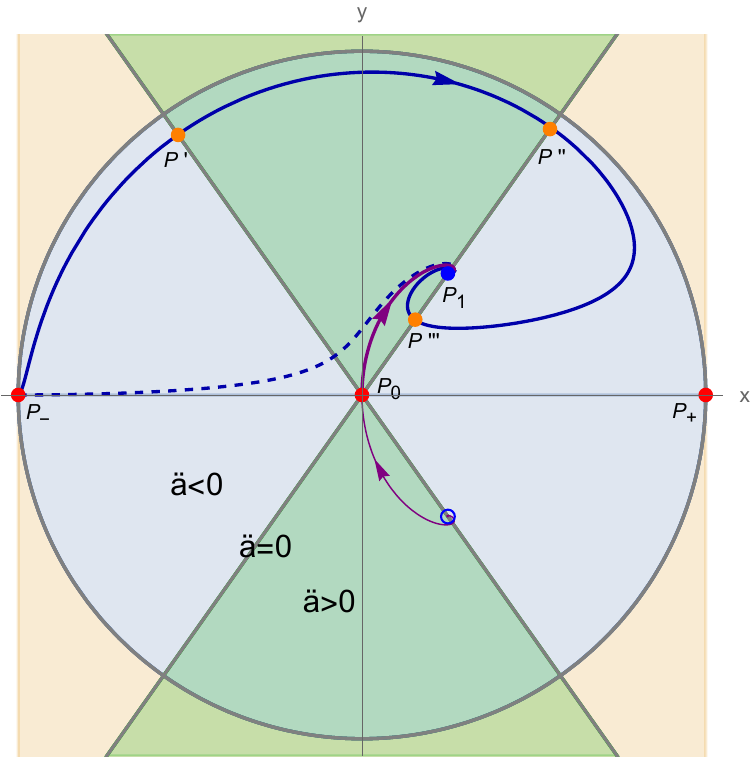}
\caption{Complete cosmological solutions asymptoting to $P_1$ as a stable spiral, displayed as trajectories in the phase space of Figure \ref{fig:PhaseSpace}, here with $d=4$ and $\gamma=\sqrt{\frac{32}{3}}$. The spiraling leads to an alternation of accelerating and decelerating phases as in a rollercoaster cosmology. In addition, the blue solutions have a ``first'' transient acceleration phase, whose number of e-folds can be increased arbitrarily by moving $P'$ towards $P_0$ as for the dashed curve. We refer to the main text for more details.}\label{fig:solsP1spiral}
\end{figure}

In Section \ref{sec:solspiral}, we have obtained solution families asymptoting to $P_1$, the latter being a stable spiral fixed point. As above, we now solve numerically the system of equations to find the complete solutions, which can asymptote to the various other fixed points indicated in Figure \ref{fig:PhaseSpace}. We display in Figure \ref{fig:solsP1spiral} these complete solutions.

Due to the spiraling, each solution asymptoting $P_1$ successively enters and exits the acceleration cone. This alternation of acceleration and deceleration phases is reminiscent of the so-called rollercoaster cosmology \cite{DAmico:2020euu}. Each cycle contributes a finite number of e-folds, computed in first approximation from \eqref{eq:ansatz3} to be
\beq
\Delta N= \frac{2\pi}{q} = \frac{\pi}{\sqrt{2}} \frac{\gamma\, \gamma_s}{\sqrt{\gamma^2 - \gamma_s^2}} \ .
\eeq

The ``initial'' phases of the solutions, closer to the other asymptotics, however differ. The solution passing through $P_0$ is special and we refer to the above discussion about it. This solution accelerates for a finite amount of time between a finite $t$ and $t=0$. As before, it admits an infinite number of e-folds between any finite time and $t=0$. The other solutions rather have a first transient acceleration phase between points $P'$ and $P''$. As before, bringing $P'$ closer to $P_0$ increases at will the number of e-folds accumulated during this first phase. This provides again to these solutions a parametric control on the number of e-folds.

\section{String theory realizations}\label{sec:string}

In this section we discuss how the cosmological solutions studied in the previous section can arise in string theory. In Subsection \ref{sec:4dstring}, we first discuss string compactifications to 4d that were studied in \cite{Marconnet:2022fmx}. We review how they give rise to consistent truncations in 4d that realize the $d=4$ theory \eqref{EFT} with $k=-1$ and with an exponential potential \eqref{V}. One example provides the fixed point $P_1$ as a stable node, and another one as a stable spiral. Both the node and the spiral cases allow for transient acceleration with parametric control of e-folds, as discussed above. The node also allows for eternal or semi-eternal acceleration. We then show in Subsection \ref{ssec:mIIA} that massive type IIA supergravity itself can match \eqref{EFT}, and give rise to $d=10$ cosmologies that are undergoing accelerated expansion for an infinite amount of time.

\subsection{4d realizations as string compactifications}\label{sec:4dstring}

We review here some results of \cite{Marconnet:2022fmx} that show that a string theory origin can be provided for certain 4d models of the form \eqref{EFT} with $k=-1$ and an exponential potential \eqref{V}.
Remarkably, as argued in Section \ref{sec:sol}, this allows for 4d cosmological solutions, realized in string theory, that are accelerating for an infinite amount of time and are accumulating an infinite number of e-folds, or
for transient acceleration with a parametric control of e-folds. This is true even though one has $\gamma > \sqrt{2}$ for the exponential rate of the potential in $d=4$, as discussed in the beginning of Section \ref{sec:sol}.

The framework of \cite{Marconnet:2022fmx} is a compactification of 10d type II supergravities with a 10d spacetime of the form (warped) $M_4\times M_6$. The 6d compact manifold $M_6$ can be either Calabi-Yau (CY), Einstein or Einstein-K\"{a}hler (EK). Reference \cite{Marconnet:2022fmx} first constructed several classes of 10d solutions, where $M_4$ is a FLRW 4d spacetime with $k=-1$ (see \eqref{eq:metric}), meaning with a negatively curved space giving an open universe. Then, a subset of these 10d solutions can equivalently be realized as 4d solutions of the following 4d theory
\eq{\label{consttr}
{\mathcal S}_{4\d}=\int\d^4 x\sqrt{g}\Big(
R-24g^{\mu\nu}\partial_\mu A\partial_\nu A
-\tfrac12 g^{\mu\nu}\partial_\mu \phi\partial_\nu\phi-V(A,\phi)
\Big)~,
}
which contains gravity and two scalar fields $A,\phi$. In other words, ${\mathcal S}_{4\d}$ is a two-scalar consistent truncation of 10d type II supergravities for cosmological solutions:~all cosmological solutions of ${\mathcal S}_{4\d}$, with a specific $V$ given below in \eqref{Vconsttrunc}, lift to 10d solutions of type II supergravities.

To see this, it is enough to show that given a certain compactification Ansatz (or a certain class of 10d solutions), the 10d equations of motion and Bianchi identities of type II supergravities boil down to exactly the equations of motion of \eqref{consttr}, with an FLRW metric: those are given by
\eq{\spl{\label{eomjp1}
\frac{\ddot{a}}{a}&=\frac16 V-8\dot{A}^2-\frac16\dot{\phi}^2\\
2\left( \frac{\dot{a}}{a} \right)^2+\frac{2k}{a^2}&=\frac13 V+8\dot{A}^2+\frac16\dot{\phi}^2\\
 \ddot{A}+3\frac{\dot{a}}{a}\dot{A}&=-\tfrac{1}{48}\partial_AV\\
 \ddot{\phi}+3\frac{\dot{a}}{a}\dot{\phi}&=-\partial_\phi V~,
}}
where  $a,A,\phi$ are assumed to depend only on time. Note that we use for now the notations of \cite{Marconnet:2022fmx}, and will eventually give a translation to conventions of the previous sections.\footnote{For the reader interested in a further comparison of conventions, let us mention that the dynamical system variables $x$, $z$ of \cite{Marconnet:2022fmx} match the $x$, $y$ of Section \ref{sec:2.2}. Also, the dynamical flow parameter $\omega$ of \cite{Marconnet:2022fmx} should be identified with the number of e-folds $N$ of \eqref{variables}. Moreover, the scale factor $a(t)$ was denoted in \cite{Marconnet:2022fmx} by $S$.}

Some cases where such a 4d consistent truncation is possible were found in\cite{Marconnet:2022fmx}\footnote{The full set of 10d solutions presented in \cite{Marconnet:2022fmx} is much richer and allows for more  fluxes to be turned on. All the  type II 10d supergravity solutions presented therein can be derived using an appropriate 1d consistent truncation with potential $U(A,B,\phi)$ (see \cite[Eq. (15)]{Marconnet:2022fmx}). This is indeed possible due to the fact that all field profiles only depend on one variable, the time. Such 1d Lagrangians can be used as a starting point for a minisuperspace treatment. The potential $V(A,\phi)$ of the 4d consistent  truncations given in \eqref{Vconsttrunc}, is related to the potential $U(A,B,\phi)$ of the 1d consistent truncations of  \cite{Marconnet:2022fmx} via
\eq{
V(A,\phi)=e^{-24A-6B}U(A,B,\phi)
~.}
However,  as already mentioned, not all the solutions of \cite{Marconnet:2022fmx} admit 4d consistent truncations. For that to be the case, the right-hand side of the equation above must be $B$-independent. This is indeed satisfied for the compactifications listed in \eqref{Vconsttrunc}.
} to be given by
\eq{
\hspace{-1.7cm}
V=\left\{\label{Vconsttrunc}
\begin{array}{ll}72\, b_0^2 \, e^{-\phi-12A}
+\frac32\, c_0^2\, e^{\frac{\phi}{2}-14A}&~  \text{CY~with~internal~3-~and~4-form~fluxes} \\ \\
\frac12\, c_{\varphi}^2 \, e^{-\frac{\phi}{2}-18A} +\frac12 \, m^2 \, e^{\frac{5\phi}{2}-6A}-6\, \lambda\, e^{-8A}&~  \text{Einstein~with external~4-form~flux} \\ \\
\frac32 \, c_0^2 \, e^{\frac{\phi}{2}-14A}+\frac12 \, m^2 \, e^{\frac{5\phi}{2}-6A}-6\, \lambda \, e^{-8A}&~  \text{EK~with internal~4-form~flux} \\ \\
\frac12\, c_{\varphi}^2 \, e^{-\frac{\phi}{2}-18A}+\frac32 \, c_f^2 \, e^{\frac{3\phi}{2}-10A}-6 \, \lambda \, e^{-8A}&~  \text{EK~with internal~2-form, external~4-form}
  \end{array}
\right.
}
where the right column describes the compactification Ansatz ($M_6$, fluxes). Those examples are all type IIA compactifications, to which we also restrict in the following. The different parameters appearing in the potential are constants coming from the fluxes of the solution, whose 10d origin is summarized in Table \ref{tab:constants}.

\begin{table}[H]
\begin{center}
\begin{tabular}{|c|c|}
  \hline
   $m$ & 0-form (Romans mass) \\
    \hline
   $c_f$ & internal 2-form  \\
     \hline
$b_0$ &internal 3-form \\
  \hline
$c_\varphi$ & external 4-form \\
  \hline
$c_0$ &internal 4-form \\
  \hline
$\lambda$ & scalar curvature of $M_6$\\
  \hline
\end{tabular}
\end{center}
\caption{List of the constant coefficients entering  the potential \eqref{Vconsttrunc} of the 4d consistent truncation, and their 10d origin as type IIA supergravity fluxes or curvature. A form is called external (internal), if all its legs are along the external 4d (internal 6d) directions.}
\label{tab:constants}
\end{table}

Given a 4d consistent truncation from 10d type IIA supergravity with $k=-1$, giving rise to the 4d theory \eqref{consttr} with scalar potential \eqref{Vconsttrunc}, one still has to get rid of one scalar field to match \eqref{EFT}. Note that the potential \eqref{Vconsttrunc} is already exponential, so one eventually reaches the scalar potential \eqref{V}. Reaching a single-scalar theory is achieved by a further consistent truncation, i.e.~a sub-truncation of \eqref{consttr}, that we will describe in two examples below.

Prior to this, let us present the Ansatz for the 10d metric, common to the two 10d solutions and their 4d realizations to be considered. The 10d metric in 10d Einstein frame reads
\eq{\label{2}\d s^2_{10} =e^{2A}\left(e^{2B} g_{\mu\nu}\d x^{\mu}\d x^{\nu}+g_{mn}\d y^m\d y^n
\right)~,
}
where the scalars $A$, $B$ are assumed to only depend on  time, while $y^m$ are coordinates of the internal 6d space.
The unwarped 4d metric  is assumed to be of the form,\footnote{On the other hand, the 4d Einstein-frame metric is given by $e^{8A+2B}(-\d \eta^2+\d\Omega_k^2)$.}
\eq{\label{metricansatz}
g_{\mu\nu}\d x^\mu\d x^\nu=
-\d \eta^2+\d\Omega_k^2
~,}
where $\eta$ is the conformal time, and the spatial 3d part of the metric is locally isometric to a maximally-symmetric 3d space of scalar curvature $6k$, namely
\eq{\label{spatialmetric}
\d\Omega_k^2=
 \gamma_{ij}(\vec{x})\,\d x^i\d x^j \ ,\quad
R^{(3)}_{ij}=2k\, \gamma_{ij}
~,}
with $i,j=1,2,3$, and $R^{(3)}_{ij}$ is the Ricci tensor of the metric $\gamma_{ij}$. In the following we will take the 3d metric to be locally that of hyperbolic space ($k<0$). We further set $k=-1$ without loss of generality, by redefining $B\rightarrow B+\text{constant}$. Note also that eventually, the degree of freedom $B$ will be traded for the scale factor $a$.

Regarding the 6d manifold $M_6$, we take it for now to be Einstein, namely
\eq{\label{150}
R_{mn}=\lambda\, g_{mn}~,
}
where $R_{mn}$ is the Ricci tensor associated to $g_{mn}$, and $6\lambda$ is the scalar curvature of $M_6$.

The rest of the Ansatz for the 10d type IIA supergravity solutions has to do with the dilaton and the fluxes. We will specify those for each of the two examples below, and show how they admit an interpretation as solutions of the 4d consistent truncation \eqref{consttr} with \eqref{Vconsttrunc}, and further as a sub-truncation to a single-scalar model \eqref{EFT} with \eqref{V}.

\subsubsection{A stable node example}

This first example has been studied in \cite{Andersson:2006du}, and more recently in \cite[Sec. 6.1]{Marconnet:2022fmx}. Considering the above 10d metric Ansatz, we further specify to a 6d Einstein manifold with negative curvature, meaning $\lambda <0$. We may set $\lambda=-1$ without loss of generality by redefining $A\rightarrow A+\text{constant}$. Furthermore, we take all the 10d fluxes to vanish. Finally, we also set the dilaton to a constant; we will see that this corresponds to a consistent sub-truncation of the 4d theory \eqref{consttr}.

This Ansatz trivially satisfies the 10d equations of motion and Bianchi identities for all the fluxes, as well as the dilaton equation of motion. The remaining equations of motion are as follows. The internal components of the Einstein
equations reduce to
\eq{\label{et3bbem}
\d_\tau^2 A=
-\lambda\, e^{16A+6B}
~,}
where we have introduced a new time variable $\tau$ defined by ${\d \eta}=e^{8A+2B}\,{\d\tau}$. The external Einstein equations reduce to the following two  equations
\eq{\spl{\label{red5bbem}
\lambda \, e^{16A+6B} -2k\, e^{16A+4B}&= \d_\tau^2B
\\
-12\lambda\, e^{16A+6B} -12k\, e^{16A+4B} &= 144\, (\d_\tau A)^2+12\, (\d_\tau B)^2 +96\, \d_\tau A\d_\tau B
~,}}
while the mixed Einstein equations (along time and internal directions) are automatically satisfied.

The three equations \eqref{et3bbem} and \eqref{red5bbem} can be ``integrated'' into the 4d action \eqref{consttr}; in other words they correspond to its equations of motion \eqref{eomjp1} with a sub-truncation. To see this, let us first introduce the cosmological (standard) time coordinate $t$ and the standard scale factor $a$ via
\eq{\label{49}
\frac{\d t}{\d\tau}={a^3}\ ,\quad a=e^{4A+B}
~.}
Taking suitable linear combinations thereof, the three
equations \eqref{et3bbem} and \eqref{red5bbem} can then be written equivalently as the 4d ones \eqref{eomjp1}, with $\phi=\text{constant}$, and $V=6|\lambda| e^{-8A}$ as in \eqref{Vconsttrunc}. Setting the dilaton to a constant is a consistent truncation of the 4d theory:\footnote{Note that this sub-truncation
is simply the restriction to the invariant plane $y=0$ of the phase space of the
corresponding dynamical system (see \cite[(91)]{Marconnet:2022fmx}). This is a special instance of a more general result: the existence of
the invariant plane~\cite[(68)]{Marconnet:2022fmx}, which therefore guarantees that all models considered in that reference
admit a consistent one-scalar sub-truncation.}
indeed, since the potential is $\phi$-independent, the equation $\ddot{\phi}+\frac{3}{a}\dot{a}\dot{\phi}=-\partial_\phi V$ is trivially satisfied. This leaves only three 4d equations to match the three 10d equations. It follows that a solution of equations \eqref{et3bbem} and \eqref{red5bbem} is also automatically a solution of the sub-truncated 4d action \eqref{consttr}.

It is then straightforward to see that this realizes the single-scalar 4d theory \eqref{EFT} with potential \eqref{V}. The 4d actions \eqref{consttr} and \eqref{EFT} then match with an overall rescaling as follows
\eq{\label{6.1}
2\sqrt{6}\, A\leftrightarrow\varphi\ ,\qquad 3 e^{-8A}\leftrightarrow V(\varphi)=3 e^{-\frac{4}{\sqrt{6}}\varphi} \ , \qquad \gamma= -\frac{V'(\varphi)}{V}=\frac{4}{\sqrt{6}}~.}
As mentioned in Section \ref{sec:sol}, this value of $\gamma$ is precisely the boundary value $\gamma_s$ for $d=4$. The fixed point $P_1$ is then a stable node (see e.g.~\eqref{P1around}), and one gets in its vicinity the accelerating solutions previously discussed.

\subsubsection{A stable spiral example}

In this second example, $M_6$ is a CY space such that the internal metric is Ricci-flat: $R_{mn}=0$. In \eqref{150}, we then take $\lambda=0$. We keep for now a non-trivial dilaton $\phi$. All fluxes are taken to vanish, except for the internal RR 2-form which is given by
\eq{\label{f2an}
  F=c_f J
~,}
where $c_f$ is a constant and $J$ is the K\"{a}hler form of $M_6$.

This Ansatz can be seen to automatically satisfy all 10d flux equations of motion and Bianchi identities. Moreover the mixed Einstein equations are automatically satisfied. The four remaining 10d equations, namely the dilaton and the internal and external Einstein equations (see \cite[(425), (426)]{Marconnet:2022fmx}), can be put in the form of the four 4d equations \eqref{eomjp1}. This is achieved with the potential $V=\frac32c_f^2\, e^{3\phi/2-10A}$, a particular case of \eqref{Vconsttrunc} with EK and $\lambda=0$ (CY), and no external 4-form. This way, the 10d solutions can also be thought of as solutions of the 4d action \eqref{consttr}.

We are left with a sub-truncation to reach a single-scalar model. With the previous potential, the right-hand side of the last two 4d equations \eqref{eomjp1} become
\beq
-\frac{1}{48} \del_A V= \frac{10}{48} \, V \ ,\quad - \del_{\phi} V = -\frac{3}{2} \, V = -\frac{36}{5} \times \frac{10}{48}\, V \ .
\eeq
It is then straightforward to see that those two equations become identical when considering\footnote{Note that this sub-truncation
is simply the restriction to the invariant plane \cite[(427)]{Marconnet:2022fmx}) of the phase space of the
corresponding dynamical system.}
\eq{
\phi=-\frac{36}{5}A~. \label{relphiA}
}
As can be checked directly from the equations of motion \eqref{eomjp1}, this choice gives a consistent sub-truncation of the action \eqref{consttr} to a single scalar $A$
\eq{\label{consttrb2}
S_{4\d}=\int\d^4 x\sqrt{g}\Big(
R-
\tfrac{1248}{25}
g^{\mu\nu}\partial_\mu A\partial_\nu A
-\frac32c_f^2\, e^{-\frac{104}{5}A}
\Big)~.
}
We then reach the single field 4d theory \eqref{EFT} with \eqref{V}. This is done with an overall rescaling and the following matching from \eqref{consttrb2}
\eq{\label{6.1b}
\frac{4}{5}\sqrt{78}A\leftrightarrow\varphi\ ,\qquad
\frac34c_f^2\, e^{-\frac{104}{5}A}
\leftrightarrow
V(\varphi)= \frac34c_f^2\, e^{-\sqrt{\frac{26}{3}}\varphi} \ ,\qquad \gamma = -\frac{V'(\varphi)}{V}= \sqrt{\frac{26}{3}}
~.}
We conclude that we get $\gamma > \gamma_s = \sqrt{\frac{8}{3}}$. As mentioned around \eqref{P1around}, the fixed point $P_1$ is then a stable spiral. This leads to 4d accelerating solutions of the kind discussed in Section \ref{sec:sol}.

\subsection{10d massive type IIA supergravity}\label{ssec:mIIA}

We now show that one of the simplest examples of accelerated cosmology of the type previously discussed is realized in massive type IIA supergravity (mIIA) in 10d \cite{Romans:1985tz}. Type IIA supergravity is a low energy limit of IIA string theory, and the mass parameter of mIIA appears naturally when  considering T-duality between type IIA and type IIB string theory \cite{Bergshoeff:2001pv}. As is well known, mIIA does not admit any straightforward covariant eleven-dimensional uplift,
see e.g.~\cite{Aharony:2010af} for a recent discussion. It is one of the two possible massive deformations of 10d IIA supergravity \cite{Tsimpis:2005vu}.
The action of massive type IIA supergravity is given in 10d Einstein frame by
\begin{align}
{\cal S}_{\rm mIIA} =& \frac{1}{2\kappa_{10}^2} \int d^{10}x \sqrt{|g_{10}|} \left( {\cal R}_{10} -\frac12 \partial_\mu \phi \partial^\mu \phi -\frac12 \left(e^{-\phi} |H_3|^2 + e^{\frac52 \phi} F_0^2+ e^{\frac32 \phi}| {F}_2|^2 + e^{\frac12 \phi}| {F}_4|^2\right) \right) \nn\\
& -\frac{1}{4\kappa_{10}^2} \int B_2 \w {F}_4\w {F}_4\,.
\end{align}
We are interested in 10d solutions whose 10d metric is the FLRW metric with $k=-1$, as given in \eqref{eq:metric}. In order to preserve the symmetry group of the maximally symmetric 9d spatial part, we set all fields except the metric, the dilaton $\phi$ and the mass parameter $F_0$ to zero.
It can readily be verified that this is indeed a consistent truncation. The action then reduces to
\begin{align}
{\cal S}_{\rm mIIA} =& \frac{1}{2\kappa_{10}^2} \int d^{10}x \sqrt{|g_{10}|} \left( {\cal R}_{10}-\frac12 \partial_\mu \phi \partial^\mu \phi -\frac12 e^{\frac52 \phi} F_0^2\right) \,,
\end{align}
which is exactly of the type \eqref{EFT} and \eqref{V} studied above. We can read off
\begin{equation}
\varphi= - \frac{1}{\sqrt{2}}\, \phi \ ,\quad   V(\phi) = \frac14\, F_0^2\, e^{- \frac{5}{\sqrt{2}} \varphi} \ ,
\end{equation}
where we set as before $M_p= \kappa_{10}^{-\frac{1}{4}} = 1 $. Note that the Bianchi identity $\d F_0=0$ gives us a constant $F_0$. So we get an exponential potential with $\gamma=\frac{5}{\sqrt{2}}$, which is larger than the lower bound $2/\sqrt{d-2}$ for $d=10$.

This setup then provides a stringy realization of the above analysis with $k=-1$. Interestingly, we recall from Section \ref{sec:sol} that $d=10$ is special, because the fixed point $P_1$ is then a stable node for any $\gamma>2/\sqrt{d-2}$. We deduce that this setup gives rise to cosmologies that undergo (semi)-eternal accelerated expansion. This is to be contrasted with the absence of de Sitter solutions in $d=10$ (see e.g.~\cite[(3.1)]{Andriot:2022xjh}).

\subsection{Accelerated expansion in the classical regime of string theory}\label{sec:classical}

We have just considered examples based on 10d type II supergravities and compactifications thereof, that admit cosmological solutions with accelerated expansion. Knowing whether those are realized in string theory requires one to verify that they are in a classical string regime, i.e.~that the supergravity approximations are justified.

Before doing so, let us recall the existence of other cosmological solutions with accelerated expansion in type II supergravity compactifications: those are de Sitter solutions, a sample of which has been obtained recently in \cite{Andriot:2022way}. Having analysed few examples of such solutions as well as considered asymptotic behaviours of related potentials \cite{Junghans:2018gdb, Banlaki:2018ayh, Andriot:2019wrs, Grimm:2019ixq, Andriot:2020vlg}, it seems however that these de Sitter solutions do not fall in a classical string regime. One reason are the numerous constraints they have to obey \cite{Andriot:2020vlg}, one being the requirement of having orientifolds (often together with $D$-branes) leading to the tadpole constraint. Having these extended objects is problematic for another reason: in the solution, one would like to take into account their backreaction. But this is a complicated problem, especially in the compactification settings considered, so the de Sitter solutions mentioned only include such objects as smeared. This refers to the fact that their contribution to the equations is taken as integrated. Although recent progress has been made on localizing these solutions \cite{Junghans:2020acz, Marchesano:2020qvg, Junghans:2023lpo}, this problem is not solved.

As we will see, the 10d supergravity solutions considered in this paper and in \cite{Marconnet:2022fmx} face none of these issues; this is a major improvement towards ensuring control and a string theory origin. First, {\sl these solutions do not have orientifolds nor $D$-branes}, so the above difficulties (and related criticisms) are avoided. Second, {\sl the classical string regime appears to be easily reached} as we now detail.

The 4d solutions discussed in Subsection \ref{sec:4dstring} have an internal volume that grows as $e^{6A}$ when $A\rightarrow \infty$. In addition, volumes of internal $p$-cycles grow as $e^{p A}$. So 4d $\alpha'$-corrections can legitimately be neglected. Turning to the string coupling constant $g_s$ governing loop corrections, we verify that it can be made small. Indeed, in the stable node case, the dilaton is fixed to an arbitrary constant that can be picked to give a small $g_s$. In the stable spiral case, the relation \eqref{relphiA} ensures that in the limit where the internal volume becomes large, $g_s$ becomes small. In other words, the time evolution goes into the direction of the classical regime.

The same holds true in the 10d realization in Subsection \ref{ssec:mIIA}. The asymptotics considered are $\phi \rightarrow - \infty$, corresponding to the classical regime. Regarding 10d $\alpha'$-corrections, one may argue that in a 10d expanding open universe, such higher derivative corrections can be neglected at a sufficiently late time. Note also that in mIIA as well as in the 4d stable spiral example, flux needs to be quantized in string theory, a point for which we do not see any difficulty.

For completeness, one could complain about the lack of control over the truncated modes in the 4d consistent truncations, which needs further investigation and is beyond the scope of this paper. One may also worry about the (non-) low energy effectiveness of the truncation (see e.g.~\cite[Sec. 5]{Andriot:2018tmb}). In the case of the CY compactification at least, this criticism may however be lifted: as the modes left after the truncation are part of a universal
consistent truncation, which is known to be a subsector of the CY effective action \cite{Terrisse:2019usq,Tsimpis:2020ysl}, we expect them to be light. However there could in general be additional light fields that are not taken into account by the truncation. Another legitimate criticism could be that we do not address moduli stabilization. At least in the massive type IIA example in $d=10$, there are no moduli to stabilize. In addition, specifying further the above $d=4$ examples, many flat field directions can be avoided. Indeed, for the stable node example, one can pick the Einstein manifold to be the 6d Poincar\'e plane divided by a discrete subgroup of SO(1,6). Due to Moscow's rigidity theorem (in other words thanks to the maximal symmetry), the resulting compact 6d manifold has only one modulus, its overall volume, which is the only one rolling. Similarly, for the stable spiral example, one can choose a rigid CY manifold. The RR 2-form flux proportional to the K\"ahler form generates a potential for all K\"ahler moduli except for the overall volume, the latter contributing again to the running field. In these examples, there is thus little concern about moduli stabilization, or at least flat directions. 

Finally, let us mention that, as can easily be seen in both these examples, the Hubble scale goes to zero at future infinity faster than the overall inverse size of the internal manifold, as measured in the 10d (and 4d) Einstein frame. So, these solutions can be viewed as (parametrically scale separated) 4d cosmologies.

Overall, the ``string theory realizations'' presented previously appear to offer a good control over possible corrections, avoiding typical problems of other constructions, thus ensuring  their string theory origin.

\section{Outlook}

In this paper, we have presented cosmological solutions describing a universe undergoing accelerated expansion, with various realizations in string theory models; our results are summarized in the Introduction. Such solutions are obtained provided the spatial curvature is negative, namely $k=-1$. One may wonder how realistic these solutions are, starting with the choice $k=-1$: we indeed recall that the minimal $\Lambda$CDM cosmological model does not consider the curvature density parameter $\Omega_k = -k/(a\, H)^2$. This is in agreement with the fact that this quantity is observationally constrained to be very close to 0  (see e.g.~\cite{ Planck:2018vyg, Bel:2022iuf}). Given its definition, $\Omega_k$ can however get very much diluted in a universe undergoing accelerated expansion: this is the standard argument in favor of inflation to solve the flatness problem. For this reason, it is very difficult to exclude the possibility of $k\neq 0$, and of having $k=-1$ as here. Nevertheless, it would be surprising to access the new physics described in this work thanks to $k=-1$, while having a negligible $\Omega_k$ compatible with $k=0$. Actually, we get an expression in terms of the phase space variables: $\Omega_k= 1- (x^2+y^2)$, i.e.~it is given by the distance to the circle with radius 1 in Figure \ref{fig:PhaseSpace}. In most solutions considered here, $\Omega_k$ is then not negligible. Regarding today's dark energy, this discussion remains meaningless without including non-relativistic matter captured by $\Omega_m$. We therefore hope to extend our analysis to include matter in future investigations.

\begin{figure}[H]
\begin{center}
\begin{subfigure}[H]{0.48\textwidth}
\includegraphics[width=\textwidth]{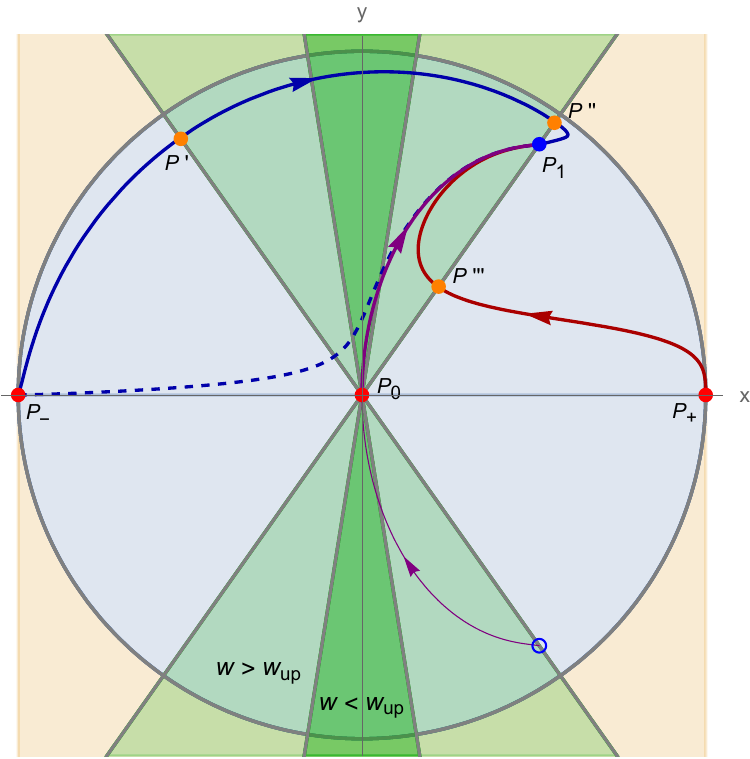}\caption{}\label{fig:wphasespace}
\end{subfigure}\quad
\begin{subfigure}[H]{0.48\textwidth}
\includegraphics[width=\textwidth]{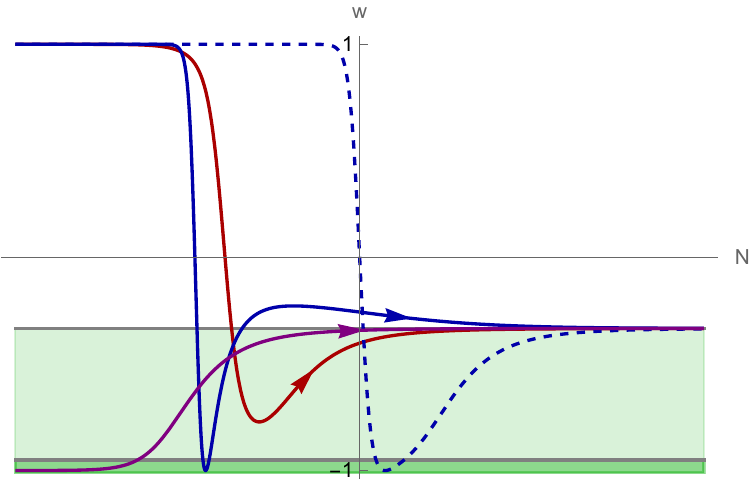}\caption{}\label{fig:wplot}
\end{subfigure}
\caption{Equation of state parameter $w$ for the cosmological solutions of Figure \ref{fig:solsP1node}. Figure \ref{fig:wphasespace} reproduces Figure \ref{fig:solsP1node} together with the observationally favored region in which $-1\leq w \leq w_{{\rm up}}$, with an upper bound $w_{{\rm up}}= -0.95$; this region is the narrower, darker green double cone. Figure \ref{fig:wplot} gives the evolution of $w$ along the cosmological solutions depicted in Figure \ref{fig:wphasespace}: the acceleration light green region  corresponds to $w < -\frac{1}{3}$, while the dark green region corresponds to $-1\leq w \leq -0.95$. Many asymptotic features of these solutions, discussed in the main text, can be identified in Figure \ref{fig:wplot}. That figure is plotted for $-10 \leq N \leq 10$.}\label{fig:w}
\end{center}
\end{figure}

A further question is how well the new cosmological solutions reproduce the acceleration observed, focusing first on the late universe. This is measured by the equation of state parameter $w$. If we consider that $\Omega_k$ is indeed negligible, we can use recent constraints on $w$. Let us take for instance from \cite{Planck:2015bue, Planck:2018vyg} the observational upper bound $w \leq w_{{\rm up}}= -0.95 $ (see also \cite{Escamilla:2023oce} for a recent account). We recall that $w=-1$ corresponds to having a cosmological constant, and theorists typically consider $w \geq -1$ to avoid scenarios with phantom energy for which no good theoretical models exist. Having $-1 < w \leq w_{{\rm up}}$ would then correspond to having a valid rolling field scenario, i.e.~a viable quintessence model. As discussed in the Introduction, the ratio $|\nabla V|/V$ or $\gamma$ can sometimes be related to $w$, but this is not enough in the complete cosmological solutions. We therefore simply obtain the value of $w$ numerically, along these solutions, and show the result in Figure \ref{fig:w}.

We first see in Figure \ref{fig:wphasespace} that the region for which $-1\leq w \leq -0.95$ is only crossed for a finite amount of time by the various solutions,\footnote{One verifies that this region corresponds to $|y| \geq |x| \, \sqrt{(1-w_{{\text up}})/(1+w_{{\text up}})}$, a double cone, provided $-1 < w_{{\text up}} < 1$.} and sometimes not at all, even though the solutions certainly go through the wider acceleration region ($w < -\frac{1}{3}$). This is made clearer in Figure \ref{fig:wplot}. It would be interesting to investigate in more detail how long this duration is, whether this can be enhanced for the various solutions and how this compares with observations.

\begin{figure}[H]
\begin{center}
\begin{subfigure}[H]{0.48\textwidth}
\includegraphics[width=\textwidth]{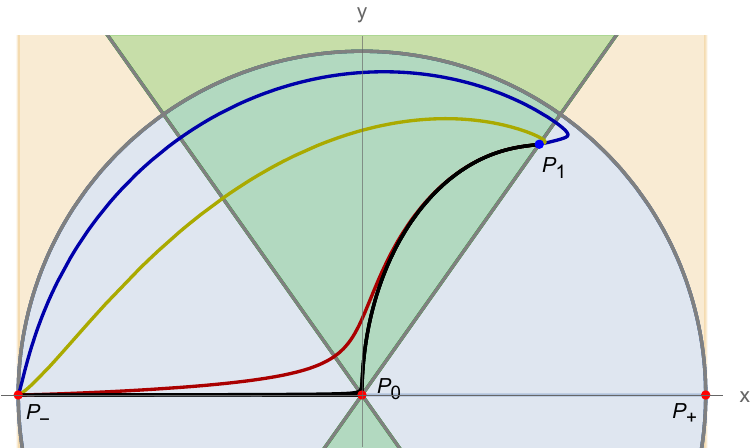}\caption{}\label{fig:wphasespaceinfnode}
\end{subfigure}\quad
\begin{subfigure}[H]{0.48\textwidth}
\includegraphics[width=\textwidth]{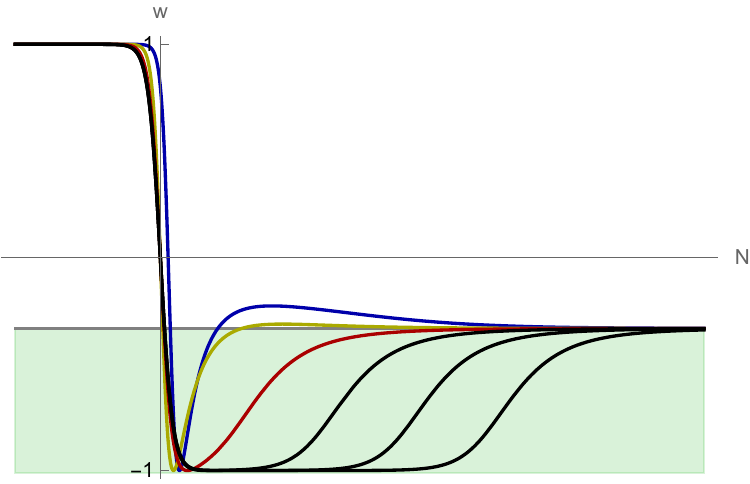}\caption{}\label{fig:wplotinfnode}
\end{subfigure}
\begin{subfigure}[H]{0.48\textwidth}
\includegraphics[width=\textwidth]{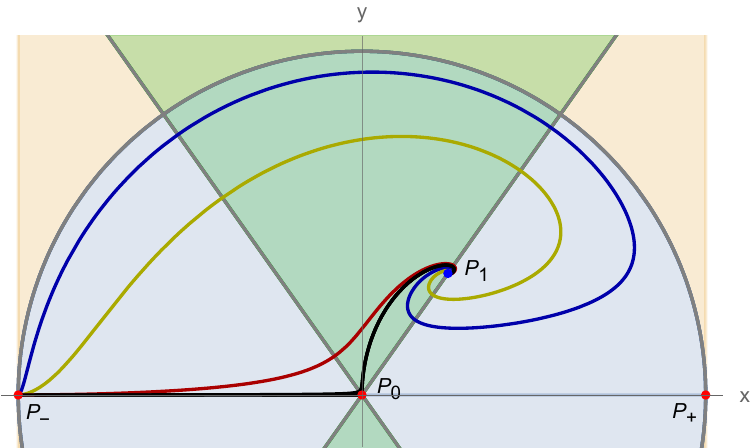}\caption{}\label{fig:wphasespaceinfspiral}
\end{subfigure}\quad
\begin{subfigure}[H]{0.48\textwidth}
\includegraphics[width=\textwidth]{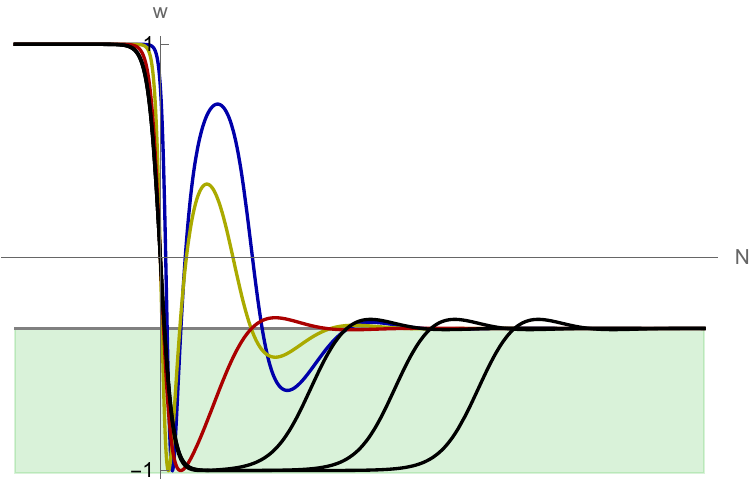}\caption{}\label{fig:wplotinfspiral}
\end{subfigure}
\caption{A realistic single-field slow-roll inflation model should admit an equation of state parameter $w \approx -1$ during a certain number of e-folds. For various solutions asymptoting to $P_1$, we show how this number of e-folds gets tuned when approaching the eternally accelerating solution, that includes a quasi-de Sitter phase. Solutions in phase space are depicted for $P_1$ being a stable node in Figure \ref{fig:wphasespaceinfnode}, resp.~a stable spiral in Figure \ref{fig:wphasespaceinfspiral}, and the corresponding $w$ are given in Figure \ref{fig:wplotinfnode}, resp.~Figure \ref{fig:wplotinfspiral}. The black solutions cannot be visually distinguished in the phase space figures. Figure \ref{fig:wplotinfnode} and \ref{fig:wplotinfspiral} are plotted for $-4 \leq N \leq 30$.}\label{fig:winf}
\end{center}
\end{figure}

Last but not least, the accelerating solutions could also be considered as models of inflation in the early universe. It would be interesting to run similar comparisons with observational constraints.  Let us recall that solutions with transient acceleration allow for a parametric control on the number of e-folds, an appealing feature when attempting to construct realistic inflation models; we illustrate this point in Figure \ref{fig:winf}. Of course not only the number of e-folds should be in agreement with constraints on inflation, but also the power spectrum predicted by the model should be consistent with observational data. Work on these
questions is in progress \cite{mtt}.

\vspace{0.4in}

\subsection*{Acknowledgements}

We would like to thank A. Hebecker, S. Parameswaran, F. Tonioni and I. Zavala for helpful discussions, as well as the organisers of the String Phenomenology 2023 Conference, that allowed these interactions. This research was made possible thanks to the support of the AAP USMB Grant dSCordes (2023). The work of T.~W.~is supported in part by the NSF grants PHY-2013988 and PHY-2210271. D.~T.~and T.~W.~would like to thank the LAPTh for hospitality; this work was initiated there.

\newpage
\providecommand{\href}[2]{#2}\begingroup\raggedright\endgroup

\end{document}